\documentclass[twocolumn]{aastex63}

\usepackage{lipsum, babel}
\usepackage{amsmath}

\newcommand{\UCI}{Department of Physics \& Astronomy, The University of California, Irvine, Irvine, CA 92697, USA}
\newcommand{\JPL}{Jet Propulsion Laboratory, California Institute of Technology, 4800 Oak Grove Drive, Pasadena, California 91109}
\newcommand{\UA}{Steward Observatory, University of Arizona, 933 N.\ Cherry Ave, Tucson, AZ 85721, USA}
\newcommand{\PSUAA}{Department of Astronomy \& Astrophysics, 525 Davey Laboratory, Penn State, University Park, PA, 16802, USA}
\newcommand{\PSUCEHW}{Center for Exoplanets and Habitable Worlds, 525 Davey Laboratory, Penn State, University Park, PA, 16802, USA}
\newcommand{\PSETI}{Penn State Extraterrestrial Intelligence Center, 525 Davey Laboratory, Penn State, University Park, PA, 16802, USA}
\newcommand{\PSUICS}{Institute for Computational and Data Sciences, Penn State, University Park, PA, 16802, USA}
\newcommand{\PSUCASt}{Center for Astrostatistics, 525 Davey Laboratory, Penn State, University Park, PA, 16802, USA}
\newcommand{\Carnegie}{Earth and Planets Laboratory, Carnegie Institution for Science, 5241 Broad Branch Road, NW, Washington, DC 20015, USA}
\newcommand{\NOAO}{NSF's National Optical-Infrared Astronomy Research Laboratory, 950 N.\ Cherry Ave., Tucson, AZ 85719, USA}
\newcommand{\TIFR}{Department of Astronomy and Astrophysics, Tata Institute of Fundamental Research, Homi Bhabha Road, Colaba, Mumbai 400005, India}

\newcommand{\Macquarie}{School of Mathematical and Physical Sciences, Macquarie University, Balaclava Road, North Ryde, NSW 2109, Australia}

\newcommand{\Carleton}{Carleton College, One North College St., Northfield, MN 55057, USA}
\newcommand{\Penn}{Department of Physics and Astronomy, University of Pennsylvania, 209 S 33rd St, Philadelphia, PA 19104, USA}

\newcommand{\GoddardESAL}{Exoplanets and Stellar Astrophysics Laboratory, NASA Goddard Space Flight Center, Greenbelt, MD 20771, USA}
\newcommand{\GoddardISTD}{Instrument Systems and Technology Division, NASA Goddard Space Flight Center, Greenbelt, MD 20771, USA}
\newcommand{\UChicago}{Department of Astronomy \& Astrophysics, University of Chicago, Chicago, IL 60637, USA}

\submitjournal{ApJ}

\begin{document}

\title{Utilizing Photometry from Multiple Sources to Mitigate Stellar Variability in Precise Radial Velocities: A Case Study of Kepler-21}

\author[0000-0001-7708-2364]{Corey Beard}
\altaffiliation{NASA FINESST Fellow}
\affiliation{\UCI}

\author[0000-0003-0149-9678]{Paul Robertson}
\affiliation{\UCI}

\author[0000-0002-0078-5288]{Mark R.~Giovinazzi}
\affil{\Penn}


\author[0000-0001-8898-8284]{Joseph M. Akana Murphy}
\altaffiliation{NSF Graduate Research Fellow}
\affiliation{Department of Astronomy and Astrophysics, University of California, Santa Cruz, CA 95064, USA}

\author[0000-0001-6545-639X]{Eric B.\ Ford}
\affil{\PSUAA}
\affil{\PSUCEHW}
\affil{\PSUICS}
\affil{\PSUCASt}

\author[0000-0003-1312-9391]{Samuel Halverson}
\affiliation{\JPL}

\author[0000-0002-7127-7643]{Te Han}
\affiliation{\UCI}

\author[0000-0002-5034-9476]{Rae Holcomb}
\affiliation{\UCI}

\author[0000-0001-8342-7736]{Jack Lubin}
\affiliation{Department of Physics \& Astronomy, University of California Los Angeles, Los Angeles, CA 90095, USA}

\author[0000-0002-4671-2957]{Rafael Luque}
\affiliation{\UChicago}

\author[0000-0001-5728-4735]{Pranav Premnath}
\affiliation{\UCI}


\author[0000-0003-4384-7220]{Chad F.\ Bender}
\affiliation{\UA}

\author[0000-0002-6096-1749]{Cullen H.\ Blake}
\affil{\Penn}

\author[0000-0002-7187-1191]{Qian Gong}
\affil{\GoddardISTD}

\author[0000-0002-0531-1073]{Howard Isaacson}
\affiliation{501 Campbell Hall, University of California at Berkeley, Berkeley, CA 94720, USA}
\affiliation{Centre for Astrophysics, University of Southern Queensland, Toowoomba, QLD, Australia}

\author[0000-0001-8401-4300]{Shubham Kanodia}
\affil{\Carnegie}

\author[0000-0001-7318-6318]{Dan Li}
\affil{\NOAO}

\author[0000-0002-9082-6337]{Andrea S.J.\ Lin}
\affil{\PSUAA}
\affil{\PSUCEHW}

\author[0000-0002-9632-9382]{Sarah E.\ Logsdon}
\affil{\NOAO}

\author[0000-0003-0790-7492]{Emily Lubar}
\affil{Aerospace Corporation, building D8, 200 N Aviation Blvd, El Segundo, CA, 90245}

\author[0000-0003-0241-8956]{Michael W.\ McElwain}
\affil{\GoddardESAL} 

\author[0000-0002-0048-2586]{Andrew Monson}
\affil{\UA}

\author[0000-0001-8720-5612]{Joe P.\ Ninan}
\affil{\TIFR}

\author[0000-0002-2488-7123]{Jayadev Rajagopal}
\affil{\NOAO}

\author[0000-0001-8127-5775]{Arpita Roy}
\affiliation{Astrophysics \& Space Institute, Schmidt Sciences, New York, NY 10011, USA}

\author[0000-0002-4046-987X]{Christian Schwab}
\affil{\Macquarie}

\author[0000-0001-7409-5688]{Gudmundur Stefansson}
\affil{Anton Pannekoek Institute for Astronomy, University of Amsterdam, Science Park 904, 1098 XH Amsterdam, The Netherlands}

\author[0000-0002-4788-8858]{Ryan C. Terrien}
\affil{\Carleton}

\author[0000-0001-6160-5888]{Jason T.\ Wright}
\affil{\PSUAA}
\affil{\PSUCEHW}
\affil{\PSETI}

\correspondingauthor{Corey Beard}
\email{ccbeard@uci.edu}

\begin{abstract}

We present a new analysis of Kepler-21, the brightest (V = 8.5) Kepler system with a known transiting exoplanet, Kepler-21 b. Kepler-21 b is a radius valley planet ($R = 1.6\pm 0.2 R_{\oplus}$) with an Earth-like composition (8.38$\pm$1.62 g/cc), though its mass and radius fall in the regime of possible ``water worlds." We utilize new Keck/HIRES and WIYN/NEID radial velocity (RV) data in conjunction with Kepler and TESS photometry to perform a detailed study of activity mitigation between photometry and RVs. We additionally refine the system parameters, and we utilize Gaia astrometry to place constraints on a long-term RV trend. Our activity analysis affirms the quality of Kepler photometry for removing correlated noise from RVs, despite its temporal distance, though we reveal some cases where TESS may be superior. Using refined orbital parameters and updated composition curves, we rule out a ``water world" scenario for Kepler-21 b, and we identify a long period super-Jupiter planetary candidate, Kepler-21 (c).

\end{abstract}

\keywords{}

\section{Introduction} \label{sec:intro}

Radial Velocities (RV) have long been an important method for discovering and characterizing exoplanets. Initial discoveries of Hot Jupiters were executed using RV instruments with precision near ten meters per second \citep{mayor95,butler96b,butler97}.  Improvements in instrument design, pipelines, and analysis methods have allowed us to detect smaller signals, such as those with lower masses and longer orbital periods \citep{mcarthur04,tinney05}. Later, the Kepler mission \citep{borucki10} would reveal that the dominant population of exoplanets consists of intermediate-sized exoplanets between the size of Earth and Neptune, dubbed Super-Earths and Sub-Neptunes \citep[e.g.,][]{howard12}. 

Measuring the masses of this common type of exoplanet is a more challenging prospect than discovering them via transits. As a transit survey, the Kepler mission could only measure planet masses for a small number of systems that exhibited transit timing variations \cite[TTVs;][]{hadden17}. RV follow-up continues to be the most reliable method for measuring planet masses, and the field continues to push RV sensitivity to smaller and smaller values, with the longstanding 1 m s$^{-1}$ noise floor recently being breached by new instruments. 

New RV instruments with on-sky (or expected on-sky) precisions well below 1 m s$^{-1}$ are available today, though they still are generally unable to characterize exoplanets with RV semi-amplitudes similar to Earth's \citep[10 cm s$^{-1}$;][]{wright17}. This is due primarily to correlated noise in the RVs from a variety of stellar astrophysics that interfere with small planetary signals, sometimes many times larger than the planetary signal in question \citep{cegla19,chaplin19, dumusque14}. These astrophysical noise sources can manifest as uncorrelated white noise, or “jitter,”  at the $\sim$ 1 m s$^{-1}$ level \citep[e.g.][]{bastien14}, or more frustratingly, they can create correlated noise that resembles false planetary signals, creating false positives \citep{lubin21}. Hence, to detect planets at the limits of our technological capacity, we must first deal with astrophysical noise sources many times larger than our instrumental variability.

High-cadence stellar lightcurves from transit surveys such as Kepler and TESS offer a powerful diagnostic tool for correcting activity contamination in RV data caused by a multitude of stellar astrophysics. For example, \cite{aigrain12} developed a diagnostic, $FF^\prime$, that is an excellent predictor of RV variability induced by rotating active regions on Sun-like stars. More recently, \cite{cegla19} predicted that high-cadence photometry may also help diagnose RV jitter from granulation at amplitudes below 1 m s$^{-1}$.

Why should stellar variability in photometry inform variability in RV data at all? Spot modulation in particular can have large effects on both datasets. Spots and plages represent regions of enhanced magnetic activity, and they show up as extra bright or dim regions in photometry. These features affect RV data by creating an anomalous red or blue shift for some small region of the star, reducing or increasing the total measured Doppler velocity \citep{dumusque14}. Spots typically rotate into and out of view, creating a periodic, or quasi-periodic, variation of both photometric brightness and measured RV. 

Importantly, while there may be phase offsets between a spot's signature in photometric or RV data, the two datasets should modulate with a related frequency, since this modulation is caused by a real, physical rotation of the star. Thus, the frequency structure of photometry and RV data should be related, if not identical.

These stellar surface features come and go, with typical lifetimes varying depending on spectral type \citep{giles17,gilbertson20}. Consequently, the frequency structure of the stellar activity will typically evolve over time, and we may not be able to enforce a strong relationship between photometry and RV data taken temporally far apart. Hence the general opinion in the field of exoplanet science that RV data taken contemporaneously with photometry is more informative for mitigating stellar activity than data taken much earlier or later \citep{haywood14}.

Temporal proximity to RV data is not the only consideration when examining photometric datasets. Kepler, for example, was a much more precise instrument than TESS (For Kepler-21: $\sigma_{\rm{Kepler, med}}$=4.38 ppm; $\sigma_{\rm{TESS, med}}$=38.38 ppm), and was sensitive to brightness variations that TESS cannot detect. Further, the observing strategy of Kepler allowed it to typically constrain periodic signals near 45 days or less \citep{mcquillan14}, while TESS's mission strategy prevents rotation estimates much greater than 12 days \citep{holcomb22}. Even if simultaneous, TESS may not be able to constrain magnetic activity in RV data if its periodicity is longer than this limit. All of these considerations have motivated us to study the tradeoffs of utilizing different photometric datasets.

We set out to study a handful of Kepler targets with the NEID spectrograph \citep{schwab16,halverson16} at Kitt Peak National Observatory (KPNO), located on the WIYN telescope\footnote{The WIYN Observatory is a joint facility of the University of Wisconsin-Madison, Indiana University, the National Optical Astronomy Observatory and the University of Missouri.},  while TESS was observing them. We were interested in studying the utility of simultaneously acquired precision RVs, and how these can best be combined with photometric data to mitigate stellar variability.

Here we present a deep dive into the RV and photometric data of one of our targets, Kepler-21, and explore the best ways to mitigate the stellar variability in its RVs using photometry. Kepler-21 was first studied by \cite{howell12}, and later in \citet[][hereafter LM16]{lopezmorales16}, and has seen extensive interest in the community for a variety of reasons. \cite{bonomo23} recently released a catalog of RV systems that further improved the mass precision of Kepler-21 b to $>$ 5$\sigma$ precision, though they did not discuss the system in depth. Kepler-21 is the brightest planet-hosting Kepler system (V = 8.5) and hosts one of the first detected exoplanets with a composition similar to Earth. Kepler-21 b also resides in the exoplanet radius valley \citep{fulton17}, which, due to the paucity of planets in the region, and our incomplete knowledge of planet formation and evolution, makes it a compelling target for atmospheric study. Despite the variety of attractive features, stellar magnetic activity (RV RMS = 5.32 m s$^{-1}$) has long challenged our ability to precisely measure the mass of Kepler-21 b, and to explore other features of the system. Utilizing a variety of new, precise RVs, Kepler-21 presents an ideal target to test the sensitivity of different activity mitigation models trained on Kepler and TESS, as well as to refine the orbital parameters of an important planetary system. We additionally place constraints on the orbital period and mass of a candidate super-Jupiter outer companion to Kepler-21 b, which we designate Kepler-21 (c).

We present an overview of the data used in our analysis in \S \ref{sec:data}. We briefly discuss our utilized stellar parameters in \S \ref{sec:stellar}. We next present an analysis of the data in \S \ref{sec:analysis}. We discuss our results in \S \ref{sec:discussion}, and provide a final summary in \S \ref{sec:summary}.

\section{Data}
\label{sec:data}

\subsection{Photometric Data}

\subsubsection{Kepler Photometry}

Kepler-21 was observed for the entire duration of the primary Kepler mission \citep{borucki10}. The Kepler spacecraft utilized a 1.4 m primary mirror to observe $>$ 190,000 main sequence stars using its 115 square degree field of view. Kepler-21 was observed from 2 May 2009 to 11 May 2013, spanning 1470 days, or just over four years. 

Kepler-21 saw both long-cadence (29.4 min) and short-cadence (58.85 s) observations. Short and long cadence data are available for all 17 Kepler quarters, with the exception of Q1, Q3, and Q4, where only long-cadence data are available. We, like LM16, utilize long-cadence data during our photometric fits, for the sake of uniformity. We utilize the Presearch Data Conditioning (PDC) flux, estimated with the Kepler Science Processing Pipeline \citep[KSPP;][]{jenkins10}. We use the \textsf{lightkurve} package (v. 2.4.1) to download Kepler PDCSAP flux data \citep{lightkurve18} for use in our transit analysis and training. The Kepler long-cadence data have an median errorbar of 4.4 ppm. Kepler data quality does vary from quarter to quarter, though not at levels that are relevant to our analysis. We utilize the \texttt{lightkurve} NaN/outlier removal functions to remove datapoints flagged as unreliable \citep{lightkurve18}.

\subsubsection{TESS Photometry}

Kepler-21 was observed by the Transiting Exoplanet Survey Satellite \citep[TESS;][]{ricker15} during Sectors 14 (18 July - 15 August 2019), Sectors 40-41 (24 June - 20 August 2021), and Sectors 53-54 (13 June - 5 August 2022). We use the presearch data conditioning simple aperture (PDCSAP) flux reduced from by the TESS science processing operations center \citep[SPOC;][]{jenkins16} pipeline during our analysis.

We again utilize built in \texttt{lightkurve} functions to remove TESS datapoints flagged as unreliable. Kepler-21 was observed both with short-cadence (2 min) and long-cadence (30 min) observations. We utilize TESS short-cadence data, which has an median error of 38 ppm. 

\subsection{Radial Velocity Data}

\subsubsection{HIRES RVs}

We utilize 49 archival and 20 newly acquired RVs of Kepler-21 taken with the High-Resolution Echelle Spectrometer (HIRES, \citealt{vogt94}), mounted on the Keck-I telescope at W.M. Keck Observatory. RVs were extracted using the iodine-cell method described in \cite{butler96a}, and we utilized archival data from \cite{butler17} for the older RVs. This archival data was run through the California Planet Search \citep[CPS;][]{howard10} pipeline in conjunction with our newly acquired HIRES RVs before analysis.

After binning observations taken on the same night, our HIRES dataset consists of 36 RVs. While more sparse than the other RV datasets, the HIRES RVs span by far the longest observation baseline, greatly expanding our sensitivity to the long period companion discussed in \S \ref{sec:trend}. HIRES observations span from 31 August 2010 to 14 July 2023. Binned HIRES RVs have an average SNR of 200, estimated as a per-pixel average at peak blaze in the middle of the iodine region (500-550 nm). This corresponds to an internal average error of 2.4 m s$^{-1}$. 

\subsubsection{HARPS-N RVs}

Our analysis utilizes 98 RV observations of Kepler-21 using the High-Accuracy Radial Velocity Planetary Searcher North \citep[HARPS-N;][]{cosentino12} located at the Telescopio Nazionale Galileo. 82 of these RVs were first published in LM16 after their analysis of Kepler-21 b, and an additional 16 were published in \cite{bonomo23}.

HARPS-N RVs have an average SNR of 167 at 550 nm, which corresponds to an average RV errorbar of 1.39 m s$^{-1}$. LM16 originally noted that this value is higher than the expected $\sim$ 1 m s$^{-1}$ uncertainty for a star of this spectral type at this SNR, and this is likely a consequence of line broadening due to the high $v \sin i$ of Kepler-21 (detailed more in \S \ref{sec:stellar}).

\subsubsection{NEID RVs}

\begin{figure}
    \centering
    \includegraphics[width=0.5\textwidth]{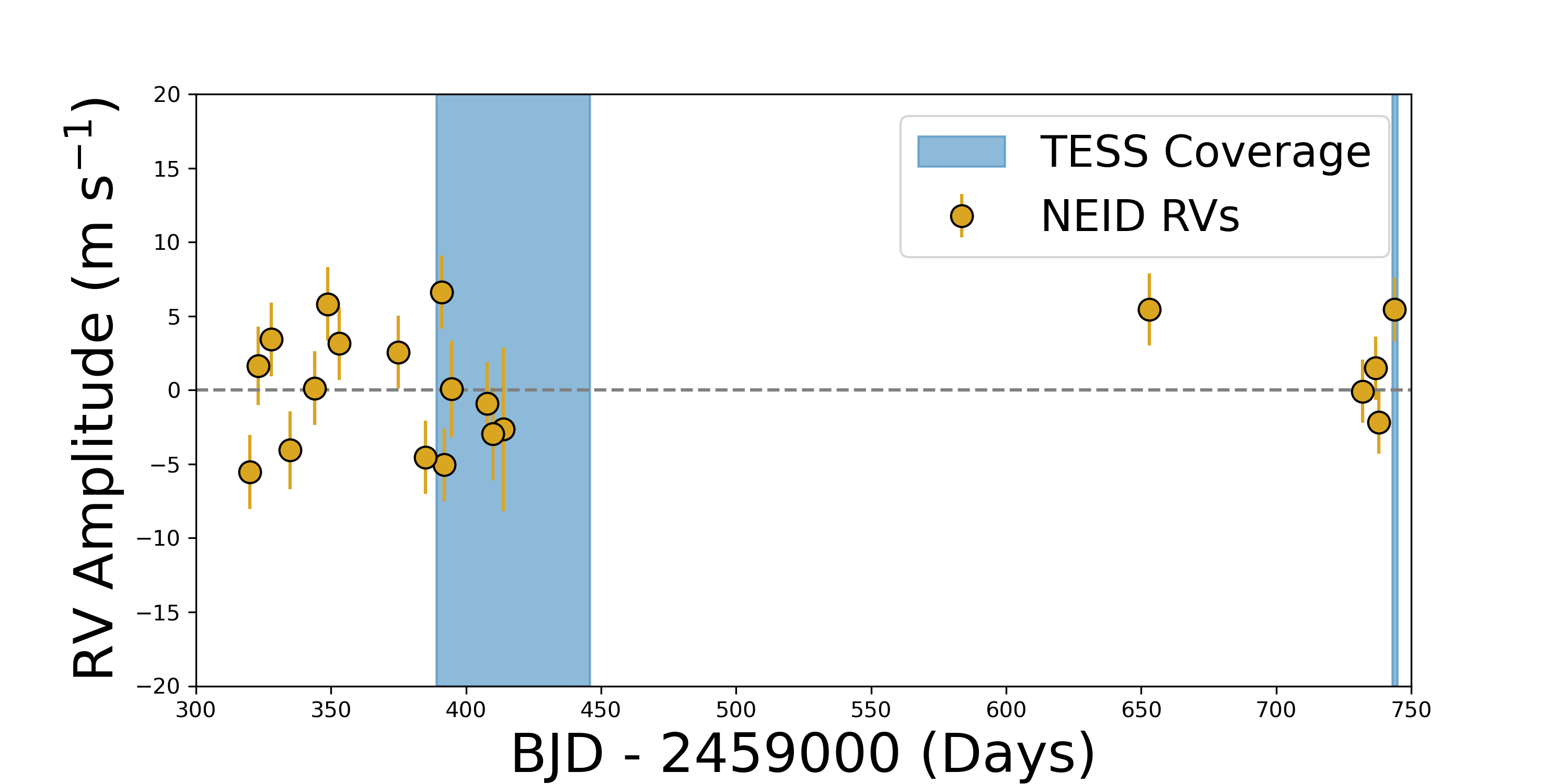}
    \caption{A plot of our NEID RV coverage as a function of time. TESS simultaneous veiwing is highlighted with a light blue streak. All of our NEID RVs were obtained near in time, or simultaneously, with TESS observations.}
    \label{fig:neid_coverage}
\end{figure}

We obtained 22 high cadence observations of Kepler-21 with the NEID spectrometer located at Kitt Peak National Observatory. NEID is an extremely stable instrument capable of obtaining RV precisions of better than 50 cm s$^{-1}$ on bright stars \citep{schwab16,halverson16}. Our observations range from 15 April 2021 to 13 June 2022, with most concentrated around TESS observation windows. Indeed, our observing strategy was to obtain as many high-cadence RVs of Kepler-21 as possible during, or near, TESS observing windows. Particularly poor weather in 2021 prevented us from obtaining as many simultaneous observations as would have been ideal, and the Contreras fire shut down Kitt Peak mere days after our observing sequence began in 2022. Despite this, we obtained seven NEID RVs simultaneous with TESS observations, with most of the rest of our RVs nearly simultaneous. Our coverage relative to TESS observations is visible in Figure \ref{fig:neid_coverage}. 

We reduced the NEID RVs using the SpEctrum Radial Velocity AnaLyser \citep[\texttt{SERVAL};][]{zechmeister18} software package, modified for use with NEID \citep[see ][]{stefansson22}. Our NEID data has a mean SNR of 102 at 490 nm, and a mean RV precision of 2.2 m s$^{-1}$.

\section{Stellar Parameters}
\label{sec:stellar}

Kepler-21 is a bright (V = 8.5), slightly evolved F5 subgiant. We follow LM16 and use the stellar parameters of Kepler-21 taken from \cite{aguirre15}, who derived stellar properties for 33 Kepler systems. Kepler-21 also saw study via asteroseismology, which allows us to constrain its age (2.84 $\pm$ 0.35 Gyr) and provides a separate measure of the system's rotation period \citep[14.83 $\pm$ 2.41 days;][]{howell12}.

Because it is a slightly evolved star with a short stellar rotation period, we expect a high $v \sin i$. LM16 measure $v \sin i = 8.4$ $\pm$ 0.5 km s$^{-1}$. This value is sufficiently high to create an effective noise floor for our most precise RVs. Indeed, the average error of our HARPS-N and NEID RVs do seem hindered by a precision limit near 1.5 m s$^{-1}$. This is especially notable for NEID, as the NEID exposure time calculator estimates a precision of 0.47 m s$^{-1}$ on a G star with a small ($<$ 2 km s$^{-1}$) $v \sin i$ of equal brightness and exposure time. Our full stellar parameters are visible in Table \ref{tab:stellar}.

\begin{deluxetable}{lllc}
\label{tab:stellar}
\tablecaption{Stellar Parameters}
\tablehead{\colhead{~~~Parameter} &
\colhead{Estimated Value} & \colhead{Units} & \colhead{Reference}
}
\startdata
 M$_{*}$ & 1.408$^{+0.021}_{-0.030}$  & M$_{\odot}$ & 1 \\
 R$_{*}$ & 1.902$^{+0.018}_{-0.012}$ & R$_{\odot}$ & 1  \\
 L$_{*}$ & 5.188$^{+0.142}_{-0.148}$ & L$_{\odot}$ & 1 \\
 $\rho_{*}$ & 0.287$^{+0.004}_{-0.005}$ & cgs & 1 \\
 $\log g_{*}$ & 4.026 $\pm$ 0.004 & cgs & 1 \\
 T$_{\rm{eff}}$ & 6305 $\pm$ 50 & K & 1  \\
 $[\rm{Fe/H}]$ & -0.03 $\pm$ 0.10  & ... & 1 \\
 Age & 2.84 $\pm$ 2.41 & Gyr & 2 \\
\enddata
\tablenotetext{ }{1 refers to \cite{aguirre15}. 2 refers to \cite{howell12}.}
\end{deluxetable}

\section{Analysis}
\label{sec:analysis}

\subsection{Training Our Models}
\label{sec:training}

Throughout the paper, and in the following sections, we will often use the terms ``Kepler-trained", or ``TESS-trained." As mentioned in \S \ref{sec:intro}, training activity models on different photometric datasets is expected to have different effects on RVs. Precision, temporal proximity, and observing baseline are expected to have the largest effect. To explore these tradeoffs, we train on Kepler and TESS photometry independently. This training consists of fitting a Gaussian Process \citep[GP;][]{ambikasaran15} to photometry prior to utilizing the posteriors of these fits as priors for RV fits.

\begin{figure}
    \centering
    \includegraphics[width=0.5\textwidth]{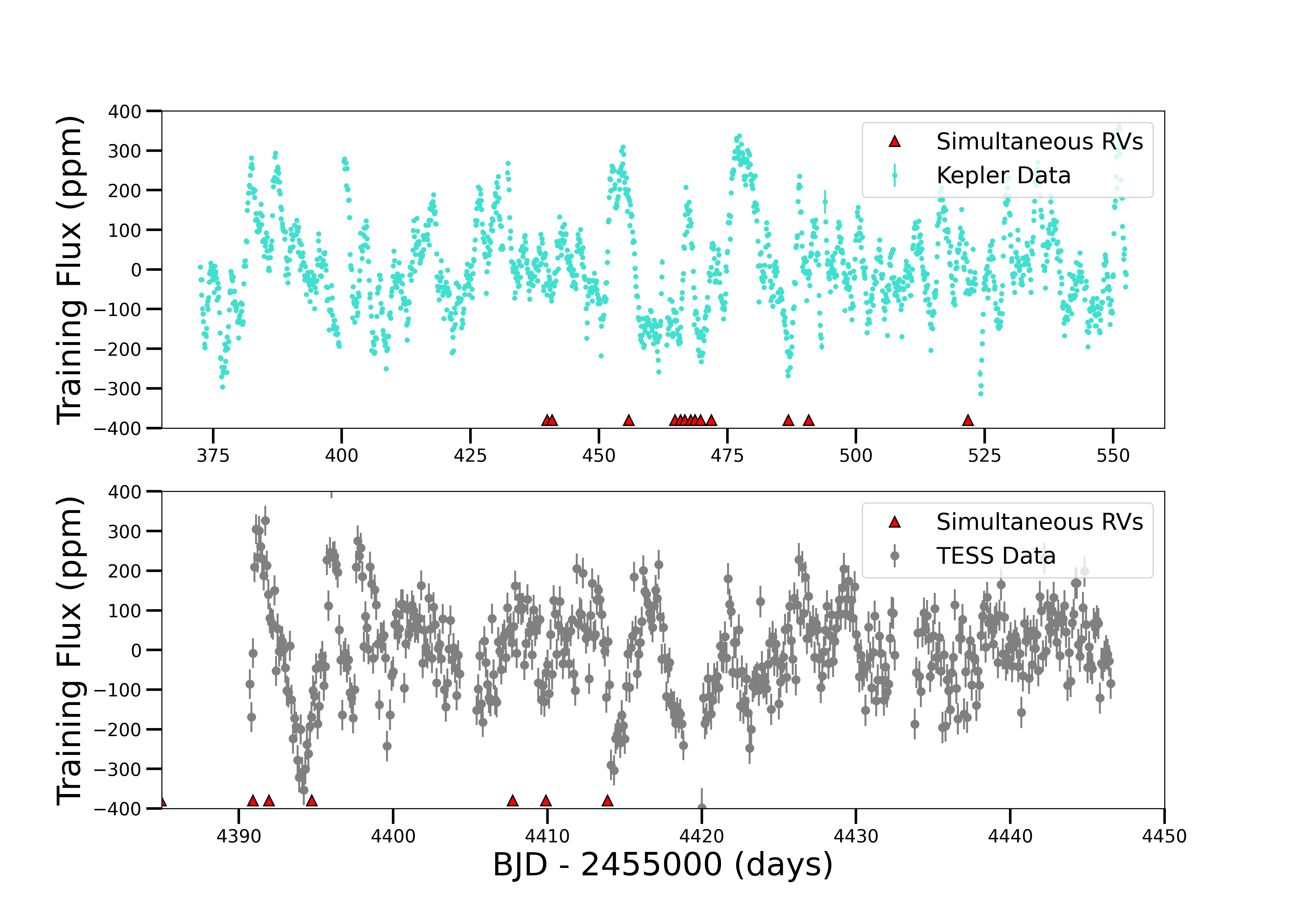}
    \caption{Top: Kepler Quarters 6 and 7 PDCSAP Flux, binned to 0.1 days. Bottom: TESS Sectors 40 and 41 PDCSAP Flux, binned to 0.1 days. Both datasets exhibit clear, quasi-periodic variability caused by stellar magnetic activity. Simultaneous RVs are marked with red triangles.}
    \label{fig:training}
\end{figure}

Because RV data is more sparse than photometry, our Kepler and TESS datasets need not be used at their full temporal resolutions. We bin our photometric data using 0.1 day bins. We also restrict our training to Sectors 40 and 41 of TESS, and Quarters 6 and 7 of Kepler. Sectors 40 and 41 were chosen because our NEID RVs were taken contemporaneously with these Sectors, and Quarters 6 and 7 of Kepler were chosen because of their simultaneity to HARPS-N RVs. Our intent is to give each dataset the best possible chance of informing the RV model most accurately, and temporal proximity is likely a deciding factor. A plot of our training data is visible in Figure \ref{fig:training}.

An important part of training a GP model is the kernel function one chooses to employ. We choose to use the $\mathcal{K}_{J1}$ GP kernel \citep{cale21}, where the i$^{\rm{th}}$ and j$^{\rm{th}}$ elements of its covariance matrix are given in equation \ref{eqn:kj1}.

\begin{equation}
    \label{eqn:kj1}
    \centering
    \footnotesize
    \mathcal{K}_{J1} = \eta_{\sigma,s(i)}\eta_{\sigma,s(j)}\exp \bigg(\frac{-|t_{i} - t_{j}|^{2}}{2\eta^{2}_{\tau}} - \frac{1}{2\eta^{2}_{l}}\sin^{2}\big(\frac{\pi|t_{i}-t_{j}|}{\eta_{p}}\big)\bigg)
\end{equation}

Above, t$_{i}$ and t$_{j}$ refer to the i$^{\rm{th}}$ and j$^{\rm{th}}$ timestamps of our timeseries, $\eta_{\tau}$ refers to the exponential decay timescale, $\eta_{l}$ is the periodic scale length, and $\eta_{p}$ is the recurrence timescale of the GP. $\eta_{\sigma,s(i)}$ refers to the amplitude hyperparameter associated with spectrograph s, utilized for observation $i$. This $\mathcal{K}_{J1}$ kernel is an expansion of the Quasi-Periodic (QP) GP kernel utilized frequently in RV exoplanet science \citep[e.g., ][LM16]{haywood14}. The QP kernel is a convenient choice not only due to its wide use, but also because of the \textit{interpretability} of its hyperparameters. $\eta_{p}$ is usually a good approximation of the stellar rotation period, $\eta_{\tau}$ can be approximated as the active region decay lifetime, and $\eta_{\sigma}$ is the amplitude of variability. $\eta_{l}$, sometimes called the structure parameter, is more difficult to interpret physically, though it is related to the number of intra-period variations the GP sees inside of a single rotation period, as LM16 explain. 

We use the $\mathcal{K}_{J1}$ kernel instead of the QP kernel because it 1) utilizes all instruments in a single covariance matrix, and 2) it utilizes a different amplitude parameter for each instrument. The former can make the model less susceptible to overfitting, which can be a serious problem when utilizing GPs on sparse datasets \citep{blunt23}. The latter is useful because the different instruments used in our analysis do not all extract RV information from the same wavelength-space, and stellar variability can be chromatic \citep{crockett12}. During photometric training, the $\mathcal{K}_{J1}$ kernel is functionally identical to the QP kernel because we only utilize one instrument at a time. The difference does become important during our RV fits in \S \ref{sec:rv_analysis}, however.

We run our training using the \texttt{RadVel} software package \citep{fulton18}. While not designed for fitting photometry, we have modified the software to use the $\mathcal{K}_{J1}$ kernel for our RV fits in \S \ref{sec:rv_analysis}, and it is trivial to evaluate a GP-only fit on a time series of any dimension. We assessed model convergence using the default method in \texttt{RadVel}, which assesses convergence by determining when the Gelman-Rubin (G-R) statistic \citep{ford06} $<$ 1.03 for all parameters, the minimum autocorrelation time factor $\geq$ 75, and a max relative change in autocorrelation time  is $\leq$ .01, and $\geq$ 1000. We adopted broad priors on the GP hyperparameters, summarized in Table \ref{tab:priors}.

The posteriors of the GP fit to Kepler and TESS photometry are used as priors in Kepler-trained and TESS-trained RV fits, respectively. The amplitude posterior from training is not used as a prior in RV fits, as its dimension is flux, not velocity, and would not make meaningful sense when applied to RV data.

Kepler-21 has a known rotation period, measured via asteroseismology to be 14.83 $\pm$ 2.41 days \citep{howell12}, and from Kepler photometry to be 12.62 $\pm$ 0.03 days (LM16). To prevent biasing our training toward Kepler, we use the asteroseismological estimate as an initial guess. Interestingly, as seen as priors in Table \ref{tab:priors}, our Kepler training produces a 22 day periodic term that we know to be erroneous. Interestingly, if we start our Kepler training at the correct 12.62 day value, the GP period term correctly settles there, even with the same priors and data. TESS training finds a 15.8 day periodicity, though with a large uncertainty. Both of these results highlight the fact that the periodic term of a GP is not strictly the same as the stellar rotation period. Rather than modify our training until we achieve a desired result, we proceed as if we do not know the true rotation period of the system. Interestingly, the Kepler-trained RV fits do recover a less precise planetary mass, as seen in Table \ref{tab:rv_posteriors}. It should be noted that our joint fits in Table \ref{tab:joint_posteriors} do recover the correct rotation period in Kepler, and its second harmonic in TESS. This might suggest a reason to prefer joint fits to RV fits trained on photometry.

\subsection{Transit Analysis}
\label{sec:transit}

Kepler-21 photometry was first obtained by the Kepler mission, with observations spanning 4.25 years. Consequently, LM16 were able to extract highly precise planetary parameters, such as orbital period and radius, by fitting only the Kepler data. 

Kepler-21 was observed with TESS for a comparatively small amount of time, only six sectors of 27 days each, adding 150 additional days of photometric data. TESS photometry is less precise than Kepler, meaning that it is unlikely to improve measured planetary parameters such as radius or impact parameter on its own. 

Utilizing both datasets together, however, is expected to improve some of the previously measured parameters. In particular, orbital period and measured time of inferior conjunction might see genuine improvements due to the significantly increased total observation baseline when including TESS data. 

In \S \ref{sec:joint}, we utilize \textit{all} Kepler quarters and TESS sectors during analysis, as we are extracting planetary parameters from the lightcurves as well. In \S \ref{sec:training} we use only the subsets described therein, as we are only extracting RV variability information.

Our transit model utilizes a mean offset for Kepler and TESS photometry and two photometric jitter terms to account for uncorrelated white noise. Kepler-21 has an estimated background contamination ratio of 8.1$\%$, suggesting that the TESS photometry is probably slightly contaminated by nearby stars \citep{stassun18}. To account for this, we scale the generated TESS light curve by the square root of a dilution term, which floats between 0 and 2 \citep{beard22}.

We use the \texttt{exoplanet}  software package \citep{exoplanet:zenodo,exoplanet:joss} to create an orbit model of Kepler-21 using its orbital period, transit time, impact parameter, stellar radius, and stellar mass. \textsf{exoplanet} utilizes this orbit model with \texttt{starry} \citep{exoplanet:luger18} to generate the light curves for Kepler and TESS, which requires a planet radius and limb darkening coefficients. We use a quadratic limb darkening law \citep{exoplanet:agol20}, and different limb darkening terms for Kepler and TESS. We use the limb darkening terms estimated from \cite{claret12, claret13} as our initial guess. 

Both Kepler and TESS photometry exhibit quasi-periodic fluctuations, likely due to stellar magnetic activity. Consequently, some model to account for this coherent noise is an essential part of modeling the photometry. We again utilize a GP to model the coherent noise, though we do not utilize the $\mathcal{K}_{J1}$ kernel described in \S \ref{sec:training} or \ref{sec:rv_analysis}. Instead, we use the \texttt{celerite2} RotationTerm, also called the double simple harmonic oscillator (dSHO) \citep{exoplanet:foremanmackey17,exoplanet:foremanmackey18}. We do so because the $\mathcal{K}_{J1}$ kernel is more computationally expensive, scaling $\mathcal{O}(N^{3})$ with number of data points, while the dSHO kernel scales as $\sim \mathcal{O}(N)$, even for large datasets. While the dSHO kernel is still widely used \citep[e.g.][]{kossakowski21,murphy23}, we find that its hyperparameters are more difficult to interpret physically. The dSHO is a combination of two simple harmonic oscillator (SHO) terms, given in equation \ref{eqn:S}.

\begin{equation}
    \label{eqn:S}
    \centering
    \rm{SHO_{n}}(\omega) = \sqrt{\frac{2}{\pi}}\frac{S_{0}\omega^{4}_{n}}{(\omega^{2} - \omega^{2}_{n})^{2} + \omega^{2}_{n}\omega^{2}/Q^{2}_{n}}
\end{equation}

In the dSHO case, our total power spectral density is the sum of SHO$_{\rm{1}}$ and SHO$_{\rm{2}}$. The actual GP fit sees a slight reparameterization using free parameters $\sigma_{GP}$, $Q_{0}$, $dQ$, $f$, and $P_{GP}$, and it restricts the frequency of the second oscillator to be twice that of the first. These are related to the above parameters by the following equations:

\begin{subequations}
    \begin{equation}
        Q_{1} = 1/2 + Q_{0} + dQ \\
    \end{equation}
    \begin{equation}
        \omega_{1} = \frac{4\pi Q_{1}}{P_{GP}\sqrt{4Q^{2}_{1} - 1}} \\
    \end{equation}
    \begin{equation}
        S_{1} = \frac{\sigma^{2}_{GP}}{(1 + f)\omega_{1}Q_{1}}
    \end{equation}
    \begin{equation}
        Q_{2} = 1/2 + Q_{0}
    \end{equation}
    \begin{equation}
        \omega_{2} = 2\omega_{1}
    \end{equation}
    \begin{equation}
        S_{2} = \frac{f}{2}S_{1}
    \end{equation}
\end{subequations}

The free hyperparameters of this kernel are $\sigma_{GP}$, the standard deviation of the process, $P_{GP}$, which is approximately the stellar rotation period, $Q_{0}$, the quality factor of the undamped harmonic oscillator, $dQ$, the difference in quality factors between SHO$_{1}$ and SHO$_{2}$, and $f$, the fractional amplitude difference of the two oscillators. While we prefer the $\mathcal{K}_{J1}$ kernel because of the more readily interpretable hyperparameters, the dSHO kernel is sufficient to model the variability in the photometry of Kepler-21.

\begin{figure*}
    \centering
    \includegraphics[width=\textwidth]{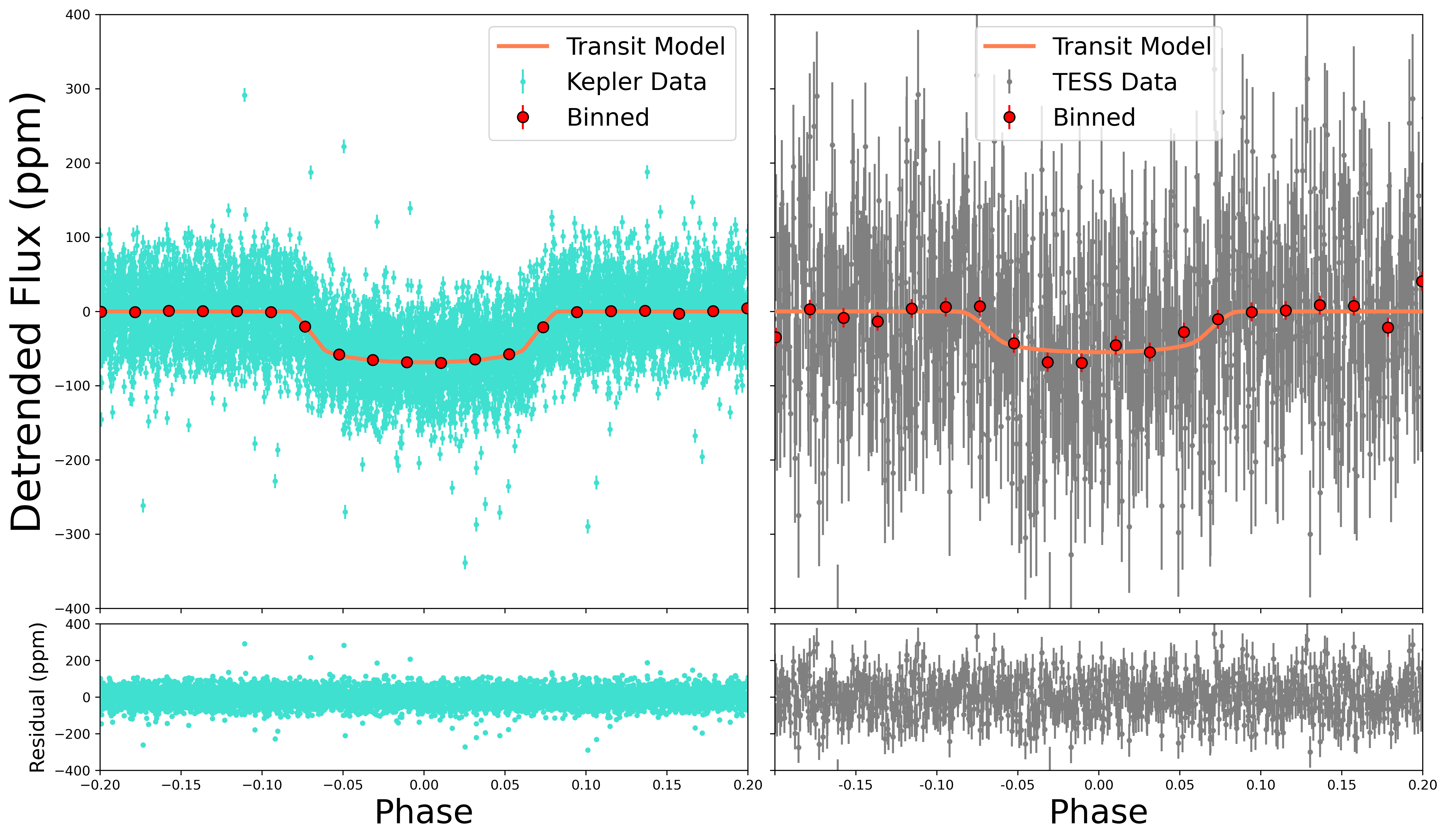}
    \caption{Left: phase folded Kepler transit of planet b, and the fit residuals below. Right: phase folded TESS transit of planet b, with residuals below. We recover the previously reported transits in Kepler, but we also include, for the first time, a fit using TESS data. The transit of a planet with a radius of 1.618 R$_{\oplus}$ orbiting a star with (R$_{*}$ = 1.902 R$_{\odot}$), is quite small, and difficult to discern from an individual transit. Nonetheless, folding multiple TESS transits and binning the data robustly reveals a transit-like structure.}
    \label{fig:transit_fit}
\end{figure*}

We generate a model context using the \texttt{PyMC3} software package \citep{exoplanet:pymc3}, which uses \texttt{theano} tensors for fast likelihood computations \citep{exoplanet:theano}. We put generally broad priors on all the free parameters of our model, with a summary in Table \ref{tab:priors}. We then explore the posterior parameter space of our model via Markov chain Monte Carlo (MCMC) inference. \texttt{exoplanet} uses the Hamiltonian Monte-Carlo (HMC) method in conjunction with a No U-Turn Sampler (NUTS) to explore the posterior parameter space efficiently \citep{hoffman11}. We run our model with two chains for 2000 tuning steps and 2000 sampling steps, confirming that our parameters are converged by ensuring their GR statistic is less than 1.001 for each. Our results are visible in Figure \ref{fig:transit_fit} and Table \ref{tab:transit}.

\begin{deluxetable}{lll}
\label{tab:transit}
\tabletypesize{\footnotesize}
\tablecaption{Transit Posteriors}
\tablehead{\colhead{~~~Parameter Name} &
\colhead{Posterior Value} & \colhead{Units}
}
\startdata
\sidehead{\textbf{Planet Parameters}}
~~~P$_{\rm{orb}}$ & 2.785823$\pm$0.000003 & Days \\
~~~T$_{\rm{c}}$ & 2459793.521$\pm$0.005 & BJD \\
~~~ e & 0.13$\pm$0.09 & - \\
~~~ $\omega$ & -10.6$\pm$45 & Degrees \\
\sidehead{\textbf{Transit Parameters}}
~~~R$_{\rm{p}}$ &  1.66$\pm$0.03 & R$_{\oplus}$ \\
~~~b &  0.64$\pm$0.08 & - \\
~~~u$_{1,Kepler}$ & 0.35$\pm$0.08 & - \\
~~~u$_{2,Kepler}$ & 0.21$\pm$0.09 & - \\
~~~u$_{1,\rm{TESS}}$ &  0.36$\pm$0.10 & - \\
~~~u$_{2,\rm{TESS}}$ &  0.23$\pm$0.10 & - \\
~~~dil$_{\rm{TESS}}$ &  0.8$\pm$0.2 & - \\
\sidehead{\textbf{Derived Parameters}}
~~~T$_{14}$  & 0.14$\pm$0.01 & Days \\
~~~a & 0.043 & AU \\
~~~i  &  88.557$\pm$0.006 & Degrees \\
~~~S  &  2752.356$\pm$0.004 & (S$_{\oplus}$) \\
~~~T$_{\rm{eq}}$ & 2015.5472$\pm$0.0008 & K \\
\sidehead{\textbf{GP Hyperparameters}}
~~~ $\sigma_{\rm{Kepler}}$ & 127$\pm$2.9 & ppm \\
~~~ $\sigma_{\rm{TESS}}$ & 129$\pm$7.4 & ppm \\
~~~ $\log Q0_{\rm{Kepler}}$ & -4.73$\pm$0.21 & - \\
~~~ $\log Q0_{\rm{TESS}}$ & -5.33$\pm$0.37 & - \\
~~~ $\log dQ_{\rm{Kepler}}$ & -1.11$\pm$0.14 & - \\
~~~ $\log dQ_{\rm{TESS}}$ & -1.10$\pm$0.70 & - \\
~~~ P$_{\rm{rot, Kepler}}$ & 5.78$\pm$0.21 & Days \\
~~~ P$_{\rm{rot, TESS}}$ & 3.48$\pm$0.48 & Days \\
~~~ f$_{\rm{Kepler}}$ & 0.016$\pm$0.002 & - \\
~~~ f$_{\rm{TESS}}$ & 0.32$\pm$0.07 & - \\
\sidehead{\textbf{Instrumental Parameters}}
~~~~~~$\gamma_{Kepler}$ & 1e6$\pm$4 & ppm \\
~~~~~~$\gamma_{\rm{TESS}}$  & 1e6$\pm$24 & ppm \\
~~~~~~$\sigma_{Kepler}$  & 50.000$\pm$0.003 & ppm \\
~~~~~~$\sigma_{\rm{TESS}}$ & 95$\pm$2 & ppm \\
\enddata
\end{deluxetable}

\begin{deluxetable*}{llllll}
\label{tab:priors}
\tabletypesize{\footnotesize}
\tablecaption{Priors Used for Various Fits}
\tablehead{\colhead{~~~Parameter Name} &
\colhead{Transit Fit Prior} & \colhead{Kepler Trained RV Fit} & \colhead{TESS Trained RV Fit} & \colhead{Joint Fit} & \colhead{Description}
}
\startdata
\sidehead{\textbf{Planet Priors}}
~~~P$_{\rm{orb}}$ (days) & $\mathcal{N}^{a}(2.785, 0.1)$ & $\mathcal{N}(2.7858212, $3e-6)  & $\mathcal{N}(2.7858212, $3e-6) & $\mathcal{N}(2.785, 0.1)$ & Orbital Period \\
~~~T$_{\rm{c}}$ (BJD) & $\mathcal{N}(2455093.8, 0.1)$ & $\mathcal{N}(2456798.7188, 0.0009) $ & $\mathcal{N}(2456798.7188, 0.0009)$ & $\mathcal{N}(2455093.8, 0.1)$ & Transit Time \\
~~~R$_{\rm{p}}$ (R$_{\oplus}$) & $\mathcal{U}^{b}(0.5, 2.0)$ & - & - & $\mathcal{U}(0.5, 2.0)$ & Radius \\
~~~K (m s$^{-1}$) & - & $\mathcal{U}(0.1, 10.0)$ & $\mathcal{U}(0.1, 10.0)$ & $\mathcal{U}(0.1, 10.0)$ & RV Amplitude \\
\sidehead{\textbf{Transit Priors}}
~~~b  & $\mathcal{U}(0.5, 1.0)$ & - & - & $\mathcal{U}(0.5, 1.0)$ & Impact Parameter \\
~~~u$_{1,Kepler}$  & $\mathcal{N}(0.3451, 0.1)$ & - & - & $\mathcal{U}(0.3451, 0.1)$ & Limb Darkening \\
~~~u$_{2,Kepler}$  & $\mathcal{N}(0.216, 0.1)$ & - & - & $\mathcal{U}(0.216, 0.1)$ & Limb Darkening \\
~~~u$_{\rm{1,TESS}}$  & $\mathcal{N}(0.3451, 0.1)$ & - & - & $\mathcal{U}(0.3451, 0.1)$ & Limb Darkening \\
~~~u$_{\rm{2,TESS}}$  & $\mathcal{N}(0.216, 0.1)$ & - & - & $\mathcal{U}(0.216, 0.1)$ & Limb Darkening \\
\sidehead{\textbf{Photometric Priors}}
~~~$\gamma_{Kepler}$  & $\mathcal{U}(0.9, 1.1)$ & - & - & $\mathcal{U}(0.9, 1.1)$ & Mean Offset \\
~~~$\gamma_{\rm{TESS}}$  & $\mathcal{U}(0.9, 1.1)$ & - & - & $\mathcal{U}(0.9, 1.1)$ & Mean Offset \\
~~~$\sigma_{Kepler}$  & $\mathcal{BLN}^{c}$(5e-5,1e-3,1e-5,7.4) & - & - & $\mathcal{BLN}$(5e-5,1e-3,1e-5,7.4) & Jitter \\
~~~$\sigma_{\rm{TESS}}$  & $\mathcal{BLN}$(5e-5,1e-3,9e-5,7.4) & - & - & $\mathcal{BLN}$(5e-5,1e-3,9e-5,7.4) & Jitter \\
\sidehead{\textbf{RV Priors}}
~~~$\gamma_{\rm{HIRES}}$  & - & $\mathcal{U}(-100, 100)$ & $\mathcal{U}(-100, 100)$ & $\mathcal{U}(-100, 100)$ & Mean Offset \\
~~~$\gamma_{\rm{HARPS-N}}$  & - & $\mathcal{U}(-100, 100)$ & $\mathcal{U}(-100, 100)$ & $\mathcal{U}(-100, 100)$ & Mean Offset \\
~~~$\gamma_{\rm{NEID}}$  & - & $\mathcal{U}(-100, 100)$ & $\mathcal{U}(-100, 100)$ & $\mathcal{U}(-100, 100)$ & Mean Offset \\
~~~$\sigma_{\rm{HIRES}}$  & - & $\mathcal{U}(0.1, 10)$ & $\mathcal{U}(0.1, 10)$ & $\mathcal{U}(0.1, 10)$ & Jitter \\
~~~$\sigma_{\rm{HARPS-N}}$  & - & $\mathcal{U}(0.1, 10)$ & $\mathcal{U}(0.1, 10)$ & $\mathcal{U}(0.1, 10)$ & Jitter \\
~~~$\sigma_{\rm{NEID}}$  & - & $\mathcal{U}(0.1, 10)$ & $\mathcal{U}(0.1, 10)$ & $\mathcal{U}(0.1, 10)$ & Jitter \\
\sidehead{\textbf{GP Priors}}
~~~$\eta_{\sigma,\rm{HIRES}}$  & - & $\mathcal{J}^{d}(0.1, 100)$ & $\mathcal{J}(0.1, 10)$ & - & GP Amplitude \\
~~~$\eta_{\sigma,\rm{HARPS-N}}$  & - & $\mathcal{J}(0.1, 100)$ & $\mathcal{J}(0.1, 10)$ & - & GP Amplitude \\
~~~$\eta_{\sigma,\rm{NEID}}$  & - & $\mathcal{J}(0.1, 100)$ & $\mathcal{J}(0.1, 10)$ & - & GP Amplitude \\
~~~$\eta_{\tau}$  & - & $\mathcal{N}(17.5, 1.6)$ & $\mathcal{N}(15.7, 1.3)$ & - & Spot Decay Timescale \\
~~~$\eta_{\rm{P}}$  & - & $\mathcal{N}(22.0, 0.1)$ & $\mathcal{N}(15.8, 2.3)$ & - & Periodic Term \\
~~~$\eta_{\rm{l}}$  & - & $\mathcal{N}(0.09, 0.03)$ & $\mathcal{N}(0.136, 0.045)$ & - & Recurrence Timescale \\
~~~$\sigma_{GP,Kepler}$  & $\mathcal{LN}^{e}$(1e-5,7.4) & - & - & $\mathcal{LN}$(1e-5,7.4) & GP Standard Deviation \\
~~~$\sigma_{GP,\rm{TESS}}$  & $\mathcal{LN}$(9e-5,7.4) & - & - & $\mathcal{LN}$(9e-5,7.4) & GP Standard Deviation \\
~~~$\sigma_{GP,\rm{HIRES}}$  & $\mathcal{U}$(0.1,100) & - & - & $\mathcal{U}$(0.1,100) & GP Standard Deviation \\
~~~$\sigma_{GP,\rm{HARPS-N}}$  & $\mathcal{U}$(0.1,100) & - & - & $\mathcal{U}$(0.1,100) & GP Standard Deviation \\
~~~$\sigma_{GP,\rm{NEID}}$  & $\mathcal{U}$(0.1,100) & - & - & $\mathcal{U}$(0.1,100) & GP Standard Deviation \\
~~~$P_{\rm{rot}}$  & $\mathcal{LU}^{f}$(1,200) & - & - & $\mathcal{LU}$(1,200) & GP Standard Deviation \\
~~~Q0  & $\mathcal{LU}$(0.002, 400) & - & - & $\mathcal{LU}$(0.002, 400) & Quality Factor \\
~~~dQ  & $\mathcal{LU}$(0.002, 400) & - & - & $\mathcal{LU}$(0.002, 400) & Difference Quality Factor \\
~~~f  & $\mathcal{U}$(0.1, 1.0) & - & - & $\mathcal{U}$(0.1, 1.0) & Fractional Amplitude \\
\enddata
\tablenotetext{a}{$\mathcal{N}$ is a normal prior with $\mathcal{N}$(mean, standard deviation)}
\tablenotetext{b}{$\mathcal{U}$ is a uniform prior with $\mathcal{U}$(lower,upper)}
\tablenotetext{c}{$\mathcal{BLN}$ is a bounded log normal prior with $\mathcal{BLN}$(lower, upper, mean, standard deviation)}
\tablenotetext{d}{$\mathcal{J}$ is a Jeffrey's prior with $\mathcal{J}$(lower,upper)}
\tablenotetext{e}{$\mathcal{LN}$ is a log normal prior with $\mathcal{LN}$(mean, standard deviation)}
\tablenotetext{f}{$\mathcal{LU}$ is a log uniform prior with $\mathcal{LU}$(lower,upper)}
\tablenotetext{-}{ indicates a free parameter that was not fit in that particular model.}
\end{deluxetable*}

\subsection{Radial Velocity Modeling}
\label{sec:rv_analysis}

We perform an RV-only analysis of Kepler-21, in addition to a joint RV + Photometry analysis in \S \ref{sec:joint}, for a number of reasons. First, we are interested in comparing the results of an RV fit when trained on Kepler or TESS data, in addition to joint fits with Kepler and TESS data. Joint fits are much more computationally expensive, so are they worth doing? Or can training on photometry first, and then fitting RV data give just as precise results? Does the precision or temporal proximity of the training photometry affect this? Most RV systems do not have the abundance of photometric data that Kepler-21 has, making joint fits of questionable value. Finally, many exoplanet systems do not transit, and it is not clear that a joint photometric fit is the right approach for such systems.

We utilize the \texttt{RadVel} software package to fit the RV data of Kepler-21. To account for stellar variability, we use the $\mathcal{K}_{J1}$ GP kernel described in \S \ref{sec:training}. 

\texttt{RadVel} models Kepler-21 b by solving Kepler's equation via a method outlined in \cite{murray99}. The orbit of each planet in the model is modified by five Keplerian parameters: orbital period (P), time of inferior conjunction (T$_{c}$), eccentricity (e), argument of periastron ($\omega$), and the RV semi-amplitude (K). Previous studies have found that Kepler-21 b has minimal eccentricity and generally fix this parameter at 0 (LM16, \cite{bonomo23}). This is generally consistent with the fact that a planet orbiting so close to an evolved star is likely to be circularized. In fact, we estimate the circularaization timescale for a planet the size of Kepler-21 b (assuming a tidal quality factor similar to Earth) as 3.5 years \citep{goldreich66}. Even with an implausible tidal quality factor one million times larger, the planet circularizes in 3.5 million years. Consequently, a nearly circular orbit is well justified and we fix eccentricity and $\omega$ to zero. 

Our model also includes a constant offset for each RV instrument, $\gamma_{inst}$, and a white noise jitter term for each instrument, $\sigma_{inst}$. This jitter term is added in quadrature with error bars when used in the likelihood computation. We also experiment with including linear and quadratic acceleration terms, as our long-baseline HIRES RVs are especially sensitive to these terms, and a preliminary analysis suggests that these may be necessary( \S \ref{sec:trend}). We run our RV fits to convergence using the same metrics in \S \ref{sec:training}. 

\begin{figure*}
    \centering
    \includegraphics[width=\textwidth]{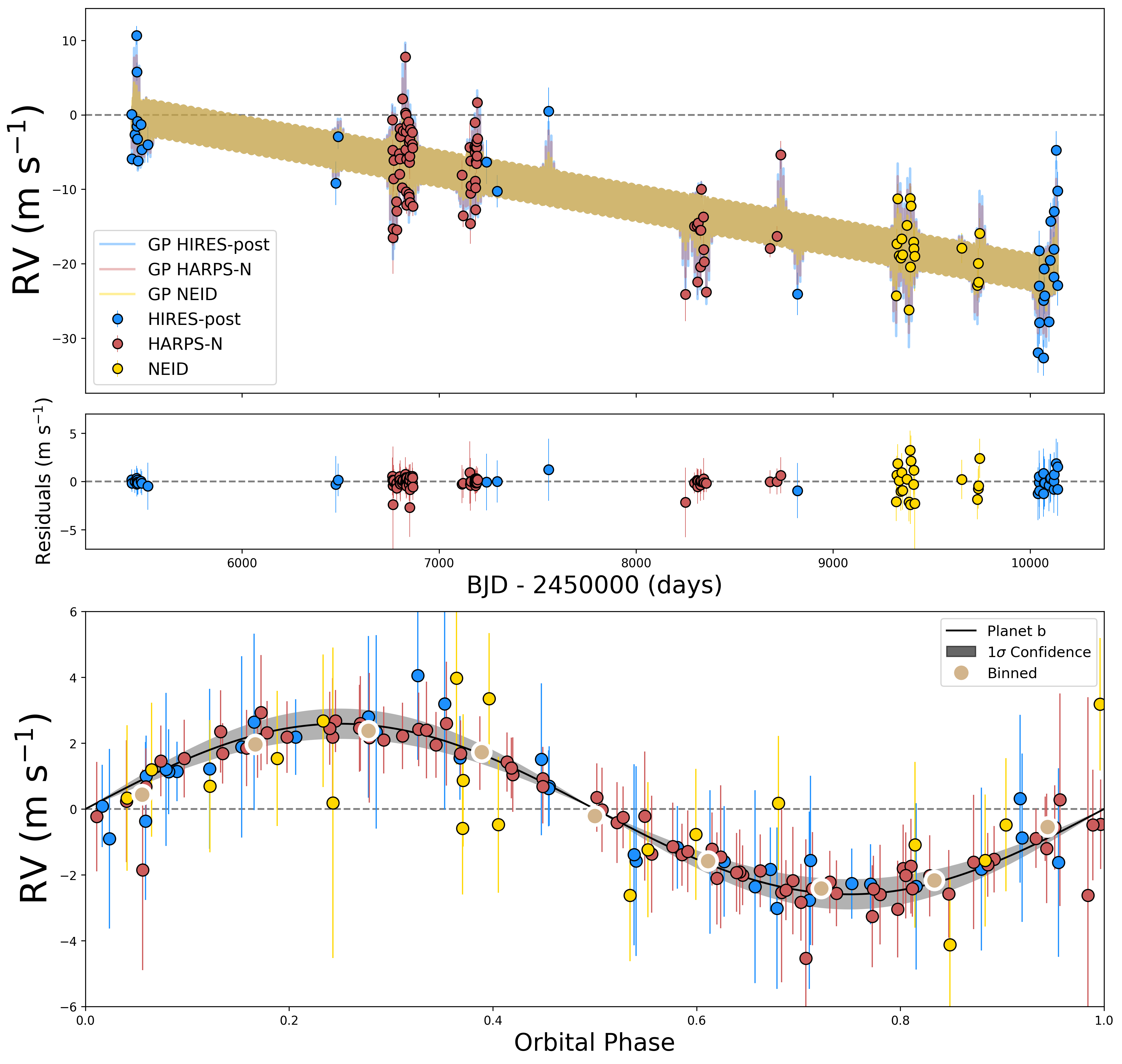}
    \caption{Top: RV time series of Kepler-21 spanning more than a decade. Light colors reflect the GP model used for each instrument. Middle: Residuals of a one planet, trend, and GP fit to the data. Bottom: phase folded RVs to planet b.}
    \label{fig:RV_fit}
\end{figure*}

We run two RV-only fits, one trained on Kepler photometry, and one trained on TESS photometry. This training changes the priors on the GP hyperparameters, but otherwise, the initial fit parameters are identical. We include a full list of priors in Table \ref{tab:priors}. The posterior estimates of our fits are detailed in Table \ref{tab:rv_posteriors}, and our best RV-only fit is visible in Figure \ref{fig:RV_fit}.

\subsection{Joint Modeling}
\label{sec:joint}

Finally, we attempt joint fits of the photometry and RVs. Such fits are the most computationally expensive option, though they can shed light where other methods do not \citep[e.g.][]{beard24}. One particular advantage is the ability to fit an activity model to photometric data and RV data simultaneously. In our RV-only fits, we train the models on the photometry, and transfer this information as informative priors. A joint fit transfers more information, as an adjustment to a shared GP parameter can immediately respond to the likelihoods of the RVs and photometry together. Planetary parameters such as orbital period and time of inferior conjunction are more strongly constrained by photometry than RV data, so we expect no difference in result due to a joint fit of these parameters. However, some GP hyperparameters can be shared between the datasets, because we expect the frequency structure of photometric and RV variability to be related. 

We use \texttt{exoplanet} to create an orbit model, and we use \texttt{starry} to generate light curves as described in our transit analysis. Due to the large quantity of data, we again utilize the dSHO kernel rather than the $\mathcal{K}_{J1}$.  Planet orbital period, time of inferior conjunction, eccentricity, argument of periastron, and orbital inclination are shared between the RV and photometric datasets. Offset and jitter terms are included for each instrument (Kepler, TESS, HIRES, HARPS-N, NEID), as described in previous sections. Our joint fits are most similar to our transit fits in \S \ref{sec:transit}, mainly due to choice of GP kernel.

The primary new feature is the treatment of GP hyperparameters between the datasets. We perform three different joint fits, which we call Kepler-RV, TESS-RV, and Kepler-TESS-RV. Each fit is performed on the same dataset utilizing all RVs and photometry. Each name emphasizes how GP hyperparameters are shared. The Kepler-RV joint fit is run such that the frequency hyperparameters of the Kepler GP are shared with the RV GP ($P_{GP}$, $Q_0$, $dQ$, $f$). We alternatively perform a TESS-RV joint fit where, again, all data are used, but the TESS GP hyperparameters (except amplitude) are shared with the RV GP. Finally, we perform a fit where all GP hyperparameters are shared between Kepler, TESS, and RV data, with the exception of an amplitude term for each. 

We use an HMC MCMC algorithm with a NUTS sampler, as described in \S \ref{sec:transit}. Due to the efficiency of the HMC, we find that our runs converge after 2000 tuning and 2000 sampling steps executed in two independent chains. Our final fit posteriors are described in Table \ref{tab:joint_posteriors}, and the priors of our joint fits are given in Table \ref{tab:priors}.

\begin{deluxetable*}{lcc}
\label{tab:rv_posteriors}
\tabletypesize{\small}
\tablecaption{RV Fit Posteriors}
\tablehead{\colhead{~~~Parameter} & \colhead{RV Fit (Kepler Trained)} & \colhead{RV Fit (TESS Trained)}
}
\startdata
\sidehead{\textbf{Orbital Parameters}}
~~~P$_{\rm{orb}}$ (days) & 2.7858212$\pm$3e-6 & 2.7858212$\pm$3e-6  \\
~~~T$_{\rm{c}}$ (BJD) & 2456798.7188$\pm$0.0008 & 2456798.7188$\pm$0.0009 \\
~~~$\sqrt{e} \cos \omega$ & - & -  \\
~~~$\sqrt{e} \sin \omega$ & - & -  \\
~~~K (m s$^{-1})$ & 2.54$\pm$0.57 & 2.55$\pm$0.47  \\
~~~$\dot{\gamma}$ (m s$^{-1}$ d$^{-1}$) & -0.0046$\pm$0.0005 & -0.0046$\pm$0.0005 \\
\sidehead{\textbf{GP Hyperparameters}}
~~~ $\eta_{\rm{\sigma,HIRES}}$ (m s$^{-1}$)& 4.8$\pm$1.8 & 5.6$\pm$1.4 \\
~~~ $\eta_{\rm{\sigma,HARPS-N}}$ (m s$^{-1}$)& 4.0$\pm$0.7 & 4.0$\pm$0.5 \\
~~~ $\eta_{\rm{\sigma,NEID}}$ (m s$^{-1}$)& 2.8$\pm$1.2 & 2.7$\pm$1.3 \\
~~~ $\eta_{\rm{\tau}}$ (days)& 16.7$\pm$1.5 & 16.0$\pm$1.3 \\
~~~ $\eta_{\rm{P}}$ (days) & 22.03$\pm$0.11 & 13.5$\pm$0.5 \\
~~~ $\eta_{\rm{l}}$ & 0.091$\pm$0.003 & 0.19$\pm$0.03 \\
\sidehead{\textbf{Instrumental Parameters}}
~~~$\gamma_{\rm{HIRES}}$ (m s$^{-1}$) & 7.8$\pm$1.7 & 7.7$\pm$1.9  \\
~~~$\gamma_{\rm{HARPS-N}}$ (m s$^{-1}$) & 6.5$\pm$1.1 & 6.3$\pm$1.3  \\
~~~$\gamma_{\rm{NEID}}$ (m s$^{-1}$) & 20.8$\pm$2.2 & 20.6$\pm$2.4  \\
~~~$\sigma_{\rm{HIRES}}$ (m s$^{-1}$) & 3.7$\pm$2.1 & 2.2$\pm$2.1 \\
~~~$\sigma_{\rm{HARPS-N}}$ (m s$^{-1}$) & 1.4$\pm$1.3 & 1.3$\pm$0.81 \\
~~~$\sigma_{\rm{NEID}}$ (m s$^{-1}$) & 2.3$\pm$1.4 & 2.6$\pm$1.4 \\
\enddata
\end{deluxetable*}

\begin{deluxetable*}{lccc}
\label{tab:joint_posteriors}
\tabletypesize{\footnotesize}
\tablecaption{Joint Fit Posteriors}
\tablehead{\colhead{~~~Parameter} & \colhead{Joint Fit (Kepler-RV)} & \colhead{Joint Fit (TESS-RV)} & \colhead{Joint Fit (Kepler-TESS-RV)}
}
\startdata
\sidehead{\textbf{Orbital Parameters}}
~~~P$_{\rm{orb}}$ (days) & 2.785823$\pm$3e-6 & 2.785823$\pm$3e-6 & 2.785823$\pm$3e-6\\
~~~T$_{\rm{c}}$ (BJD) & 2455093.8364$\pm$0.0009 & 2455093.8364$\pm$0.0009 & 2455093.8365$\pm$0.0009 \\
~~~$\sqrt{e} \cos \omega$ & - & - & - \\
~~~$\sqrt{e} \sin \omega$ & - & - & - \\
~~~K (m s$^{-1})$ & 2.48$\pm$0.48 & 2.66$\pm$0.57 & 2.41$\pm$0.49 \\
~~~$\dot{\gamma}$ (m s$^{-1}$ d$^{-1}$) & -0.0049$\pm$0.0005 & -0.0047$\pm$0.0005 & -0.0046$\pm$0.0005 \\
\sidehead{\textbf{Transit Parameters}}
~~~R$_{\rm{p}}$ (R$_{\oplus}$) &  1.65$\pm$0.02 & 1.65$\pm$0.02 & 1.65$\pm$0.02 \\
~~~b  & 0.620$\pm$0.014 & 0.620$\pm$0.014 & 0.620$\pm$0.014 \\
~~~u$_{1,Kepler}$ &  0.35$\pm$0.07 & 0.35$\pm$0.08 & 0.34$\pm$0.08 \\
~~~u$_{2,Kepler}$ & 0.21$\pm$0.09 & 0.21$\pm$0.09 & 0.21$\pm$0.09 \\
~~~u$_{1,\rm{TESS}}$ &  0.36$\pm$0.10 & 0.36$\pm$0.10 & 0.36$\pm$0.10 \\
~~~u$_{2,\rm{TESS}}$ & 0.23$\pm$0.10 & 0.23$\pm$0.10 & 0.23$\pm$0.10 \\
~~~dil$_{\rm{TESS}}$ &  0.80$\pm$0.18 & 0.80$\pm$0.18 & 0.75$\pm$0.15 \\
\sidehead{\textbf{Derived Parameters}}
~~~M$_{\rm{p}}$ (M$_{\oplus}$) & 6.9$\pm$1.3 & 7.4$\pm$1.6 & 6.68$\pm$1.4 \\
~~~T$_{14}$ (days) & 0.144$\pm$0.002 & 0.1442$\pm$0.002 & 0.144$\pm$0.002 \\
~~~a (AU) & 0.0434281$\pm$3e-8 & 0.0434281$\pm$3e-8 & 0.0434281$\pm$3e-8 \\
~~~i (degrees) & 88.556$\pm$0.003 & 88.556$\pm$0.003 & 88.556$\pm$0.003 \\
~~~S (S$_{\oplus}$) &  2752.356$\pm$0.004 & 2752.356$\pm$0.004 & 2752.356$\pm$0.004 \\
~~~T$_{\rm{eq}}$ (K) &  2015.5472$\pm$0.0008 & 2015.5472$\pm$0.0008  & 2015.5473$\pm$0.0008\\
~~~$\rho$ (g/cc) & 8.92$\pm$1.96 & 9.18$\pm$1.86 & 8.38$\pm$1.62\\
\sidehead{\textbf{GP Hyperparameters}}
~~~ $\sigma_{\rm{GP, Kepler}}$ (ppm) & 151$\pm$5 & 127$\pm$3 &  145$\pm$4 \\
~~~ $\sigma_{\rm{GP, TESS}}$ (ppm) & 129$\pm$7 & 131$\pm$7 & 278$\pm$2 \\
~~~ $\sigma_{\rm{GP, HIRES}}$ (m s$^{-1}$) & 6.33$\pm$1.0 & 6.0$\pm$0.9 & 7.5$\pm$1.3  \\
~~~ $\sigma_{\rm{GP, HARPS-N}}$ (m s$^{-1}$) & 4.2$\pm$0.5 & 4.1$\pm$0.5 & 4.9$\pm$0.7 \\
~~~ $\sigma_{\rm{GP, NEID}}$ (m s$^{-1}$) & 3.3$\pm$1.2 & 3.4$\pm$1.2 & 3.3$\pm$1.6  \\
~~~ $\log Q0_{\rm{Kepler}}$ & -4.8$\pm$0.21 & -4.73$\pm$0.14 & -2.31$\pm$0.07 \\
~~~ $\log Q0_{\rm{TESS}}$ & -5.3$\pm$0.4 & -5.6$\pm$0.3 & -2.31$\pm$0.07 \\
~~~ $\log dQ_{\rm{Kepler}}$ & -1.1$\pm$0.1 & -1.1$\pm$0.3 & 3.35$\pm$0.30 \\
~~~ $\log dQ_{\rm{TESS}}$ & -1.1$\pm$0.7 & -1.6$\pm$0.6 & 3.35$\pm$0.30 \\
~~~ P$_{\rm{rot, Kepler}}$ (days) & 5.8$\pm$0.2 & 5.77$\pm$0.22 & 12.53$\pm$0.14  \\
~~~ P$_{\rm{rot, TESS}}$  (days) & 3.5$\pm$0.5 & 4.01$\pm$0.61 & 12.53$\pm$0.14 \\
~~~ f$_{\rm{Kepler}}$ & 0.016$\pm$0.002 & 0.015$\pm$0.002 & 0.94$\pm$0.05 \\
~~~ f$_{\rm{TESS}}$ & 0.32$\pm$0.7 & 0.30$\pm$0.06 & 0.94$\pm$0.05 \\
\sidehead{\textbf{Instrumental Parameters}}
~~~\underline{Photometric} \\
~~~~~~$\gamma_{Kepler}$ (ppm) & 1e6$\pm$4 & 1e6$\pm$4 & 1e6$\pm$4 \\
~~~~~~$\gamma_{\rm{TESS}}$ (ppm) & 1e6$\pm$10 & 1e6$\pm$12 & 1e6$\pm$24 \\
~~~~~~$\sigma_{Kepler}$ (ppm) & 119.256$\pm$0.007 & 50.003$\pm$0.003 & 50.000$\pm$0.003 \\
~~~~~~$\sigma_{\rm{TESS}}$ (ppm) & 82$\pm$2 & 83$\pm$2 & 95$\pm$2\\
~~~\underline{RV} \\
~~~~~~$\gamma_{\rm{HIRES}}$ (m s$^{-1}$) & -3.2$\pm$1.2 & -3.2$\pm$1.1 & -3.13$\pm$1.2 \\
~~~~~~$\gamma_{\rm{HARPS-N}}$ (m s$^{-1}$) & -4.7$\pm$0.7 & -4.6$\pm$0.6 & -4.6$\pm$0.7 \\
~~~~~~$\gamma_{\rm{NEID}}$ (m s$^{-1}$) & 9.9$\pm$1.3 & 9.8$\pm$1.2 & 9.7$\pm$1.2 \\
~~~~~~$\sigma_{\rm{HIRES}}$ (m s$^{-1}$) & 5.0$\pm$2.9 & 5.0$\pm$2.9 & 5.7$\pm$2.9 \\
~~~~~~$\sigma_{\rm{HARPS-N}}$ (m s$^{-1}$) & 3.6$\pm$2.4 & 4.1$\pm$2.8 & 3.8$\pm$2.6 \\
~~~~~~$\sigma_{\rm{NEID}}$ (m s$^{-1}$) & 4.9$\pm$2.9 & 5.5$\pm$2.9 & 5.1$\pm$2.5 \\
\enddata
\end{deluxetable*}

\subsection{Injection Recovery Analysis}
\label{sec:injection}

We perform injection-recovery tests to explore the value of photometric training datasets. Kepler photometry is more precise than TESS, and its longer baseline provides many advantages. It is also temporally separated from newer RV data by more than ten years. This is particularly concerning given that stellar variability is known to evolve on much shorter time scales \citep{giles17,gilbertson20}. We can begin to evaluate the training datasets by noting which produced the most precise orbital parameters in the previous section, though comparing recovered mass precisions is not necessarily the best way to determine the best activity mitigation method. In particular, such a method could not be expanded to systems with no known planets, despite interest in knowing which photometric dataset would be best for \textit{searching} for planets in such systems. 

We follow a method similar to \cite{cale21} by injecting circular planet signals into the data at a variety of orbital periods and RV amplitudes. We choose ten RV amplitude bins ranging between 0 to 10 m s$^{-1}$, uniformly spaced. For orbital periods, we create ten bins between 1 and 1000 days with log-uniform spacing. We then inject signals randomly drawn from within these two dimensional bins into our data, and fit both a one- and two-planet model. Targets are also injected with a random time of periastron, drawn from a distribution with a width equal to the injected period. Especially for longer period planets, this is an important step to take to simulate the effects of favorable or unfavorable phase coverage. This process is executed with models trained on Kepler and TESS, as well as on ``untrained" datasets where broad, uninformative GP hyperparameter priors are used. Each fake signal is added to the existing Kepler-21 RV data to create a fake dataset, and we do this 100 times in each period-K bin.

We use the Bayes Factor \citep[BF;][]{kass95} for model comparison to determine if an injected planet is recovered successfully. The BF is produced by taking the ratio of the evidence of two competing models, with a value $>$ 1 indicating preference for the numerator. Typically, we require that a more complicated model be substantially better than a simpler model to warrant adoption. This threshold is not important during our injection-recovery tests where we are interested only in comparing which training method recovers a higher BF, though in \S \ref{sec:trend} we do require a log$_{10}$(BF) $>$ 5 to prefer a more complicated model, which is a commonly adopted threshold \citep[i.e.][]{trotta08,luque19, beard22}.

\begin{figure*}
    \centering
    \includegraphics[width=\textwidth]{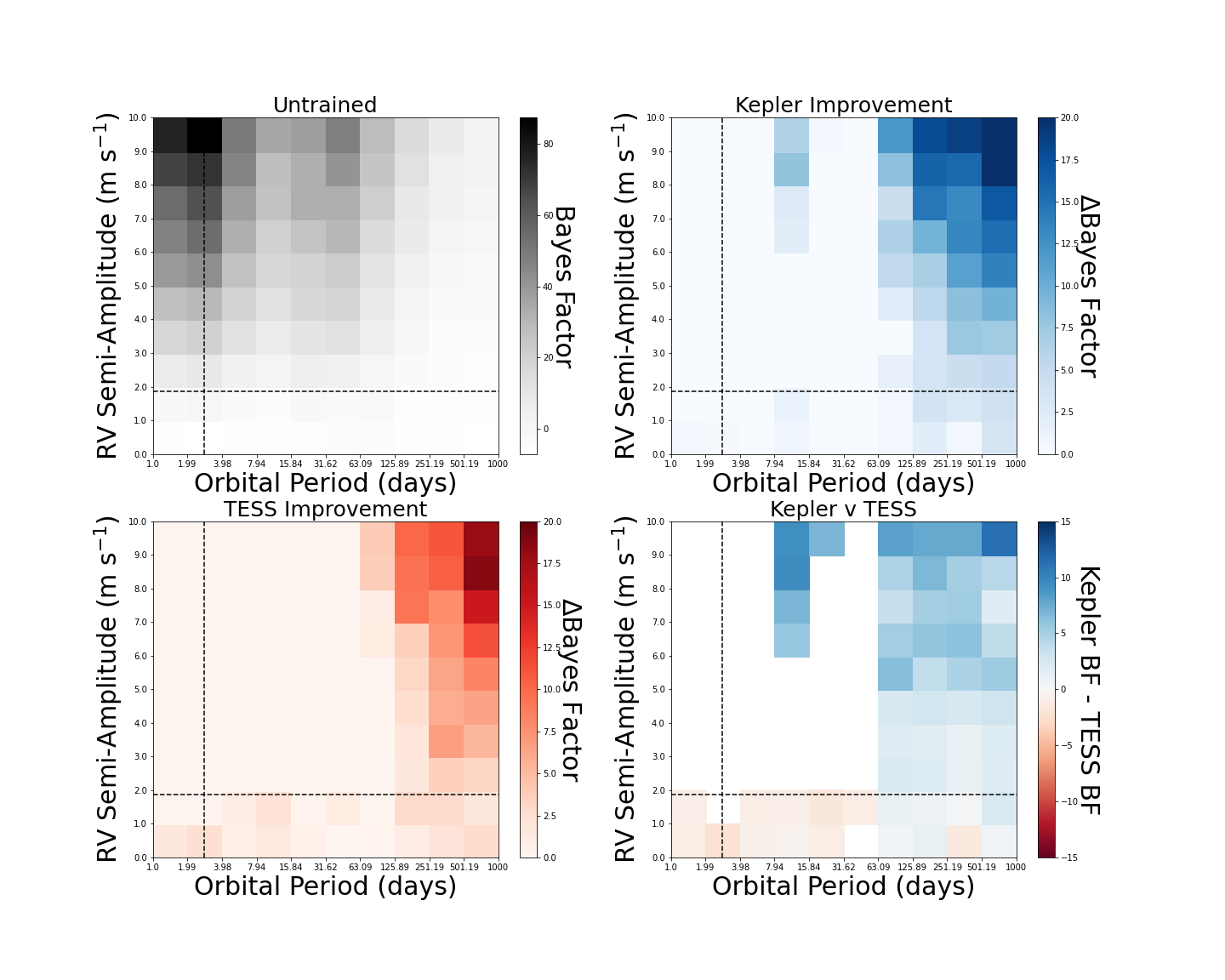}
    \caption{Injection recovery results for fits with an injected “transiting" planet. The amplitude and period of the known planet are denoted by a black dashed line. Top Left: Injection recovery tests run with no GP training.  Top Right: differential preference for recovering the injected planets between Kepler and untrained fits. Bottom Left: differential preference for recovering the injected planets between TESS and untrained fits.  Bottom Right: differential BF improvements between the training methods. Longer orbital periods are consistently recovered more robustly when training on Kepler.}
    \label{fig:IR_fixed}
\end{figure*}

\begin{figure*}
    \centering
    \includegraphics[width=\textwidth]{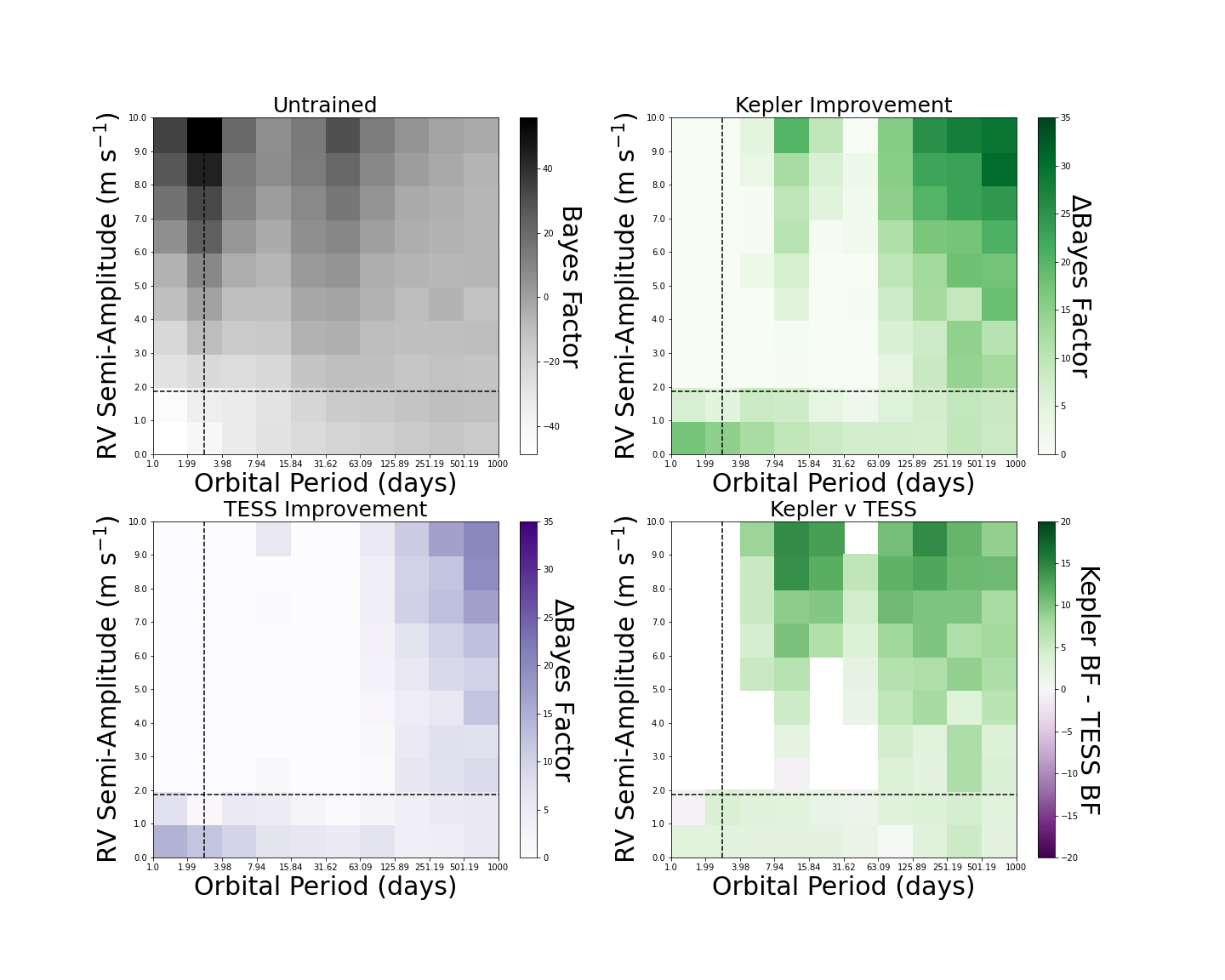}
    \caption{Injection recovery results for fits with an injected “RV-detected" planet. The amplitude and period of the known planet are denoted by a black dashed line. Top Left: Injection recovery tests run with no GP training.  Top Right: differential preference for recovering the injected planets between Kepler and untrained fits. Bottom Left: differential preference for recovering the injected planets between TESS and untrained fits.  Bottom Right: differential BF improvements between the training methods. The vast majority of injected planets were recovered more strongly when training on Kepler.}
    \label{fig:IR_vary}
\end{figure*}

Continuing to follow \cite{cale21}, we explore two different cases of injected planet. First, we explore the case of a “transiting" injected planet, where the orbital period of the injected planet is known, and we fix it to the correct, injected value. Second, we explore the “RV detected" injected planet, where the period and time of inferior conjunction of the injected planet is not known a priori. The results of these two tests are visible in Figures \ref{fig:IR_fixed} and \ref{fig:IR_vary}.

We generate our injected signals using the \texttt{RadVel} software package. Because of the large number of RV fits required for our analysis, we make a few simplifying assumptions during injection-recovery. In contrast to the RV fits in \S \ref{sec:rv_analysis}, we fix the orbital period and time of conjunction of Kepler-21 b. We only inject circular planets, and always fix our model eccentricity at zero. An exploration of how our results are sensitive to eccentricity would be interesting, but is beyond the scope of our work. We use the injected orbital periods and RV semi-amplitudes as starting guesses for all of our two planet models. In the case of “RV detected" planets, we place generally broad priors on the orbital period and time of inferior conjunction of the second planet. The orbital period has a Gaussian prior centered at P$_{inj}$ with a standard deviation of P$_{inj}$/2. The time of inferior conjunction has a uniform prior centered at the injected value with a width of the injected period.

With a 10$\times$10 grid of bins, and 100 signals injected in each bin, we perform 10,000 signal injections for the Kepler-trained models and TESS-trained models respectively, as well as an untrained RV fit. Since we are fitting both a one and two planet model to each injected signal, we perform two RV fits per signal, resulting in 20,000 RV fits for Kepler-trained, TESS-trained, and untrained models, or 60,000 total. Performing these analyses on the ``transiting" and ``RV-detected" cases brings our final number of RV fits to 120,000.

Performing an MCMC or nested sampling fit to each model would be prohibitively expensive in terms of CPU-hours, and so we approximate the evidence of each model using the Laplace approximation, given in equation \ref{eqn:laplace} \citep{nelson20}. The Laplace approximation allows us to estimate the evidence of each model fit, rather than integrating it numerically. The evidence is defined in equation \ref{eqn:evidence}, \citep{kass95}.

\begin{equation}
    \label{eqn:evidence}
    \centering
    \mathcal{E} = \int \mathcal{L}(d,\theta)\pi(d,\theta)d\theta
\end{equation}

Here, $\mathcal{E}$ is the evidence of a model, $\mathcal{L}(\theta)$ is the likelihood of the model with data $d$ and parameters $\theta$. $\pi(\theta)$ is the prior probability of the parameters. This integral is typically intractable to calculate, and the Laplace approximation circumnavigates this computation. It is convenient to rewrite the right term $\mathcal{L}$ as an exponentiated logarithm, exp$(\log (\mathcal{L}(d,\theta)\pi(\theta)))$. Then we can say that it is equal to

\begin{equation}
    \label{eqn:laplace}
    \centering
    \bigg[\frac{(2\pi)^{2}}{|\det(H(d,\theta_{0})|}\bigg]^{1/2} \exp(\log(\mathcal{L}(d,\theta_{0})\pi(\theta_{0})))
\end{equation}

where H is the Hessian matrix of the function $\log(\mathcal{L}(d,\theta)\pi(\theta))$, and $\theta_{0}$ is a set of parameters that produce a local maximum. This approximation is generally true for a function where the dominant mode is well separated from the integration domain, which is true for our analysis. We first optimize our initial values using a least squares fit in \texttt{scipy.optimize.minimize}, and then we estimate the evidence at this point. We take the average $\Delta$BF in each bin and report them in Figures \ref{fig:IR_fixed} and \ref{fig:IR_vary}.

\subsection{Additional Bodies in the System?}
\label{sec:trend}

The Kepler spacecraft observed Kepler-21 for more than four years, and only identified transits of Kepler-21 b. It is not feasible that TESS would observe two or more planetary transits that Kepler missed, but it is possible that a single, long-period planet transits in TESS, but not Kepler. We use the \textsf{Transit Least Squares} \citep[TLS;][]{hippke19} Python package to search for additional transits in Kepler-21. Due to the correlated noise present in the photometry, we applied a UnivariateSpline to the Kepler and TESS photometry to remove the noise while minimally affecting the transits \citep{dierckx95}. After recovering and masking the transits of Kepler-21 b, we run the TLS algorithm on the collective photometry, but we detect no significant additional transit events. We conclude that there are no additional observed transiting planets in Kepler-21 up to the limits of our photometric sensitivity.

Our analysis adds to the already significant ($\sim$ 10 years) RV baseline of the system, extending our sensitivity to longer period, non-transiting planets. The observing baseline of HIRES allows us to constrain the RV offsets present in other instruments. The data seem strongly suggestive of either a trend or curvature, and we perform a series of RV fits to determine which explanation is best. In Table \ref{tab:model_comparison}, we compare a variety of models exploring no trend, a linear trend, a fit with a linear and quadratic term, and a two planet model. We use the Bayesian Information Criterion \citep[BIC;][]{kass95, liddle07} to approximate the BF between our models. Although a TESS-trained linear + quadratic model has the lowest BIC value, it is not low enough to justify adopting the more complicated model. The linear trend fits are significantly better than the no-trend cases, however, and so we adopt a linear trend. Our best fits recover a $\dot{\gamma}$ value of -0.0046$\pm$0.0005 m s$^{-1}$ day$^{-1}$, which is 9.2$\sigma$ significant. This could be caused by a long period activity cycle, though analysis of the HIRES S$_{HK}$ values reveals no such corresponding trend. We conclude that this trend is suggestive of either an additional, long-period planet, or perhaps a substellar companion.

To identify the source of the linear RV trend, we searched the Gaia database for possible bound stellar companions. Gaia reports Kepler-21 to have a renormalized unit weight error \citep[RUWE;][]{lindegren18} consistent with unity, a strong indicator that Kepler-21 is not an unresolved stellar multiple. We do not find any co-moving stars with similar parallaxes in Gaia \citep{luri18}, suggesting that the cause of this apparent RV slope is the result of a substellar object. 

\begin{figure}
    \centering
    \includegraphics[width=0.5\textwidth]{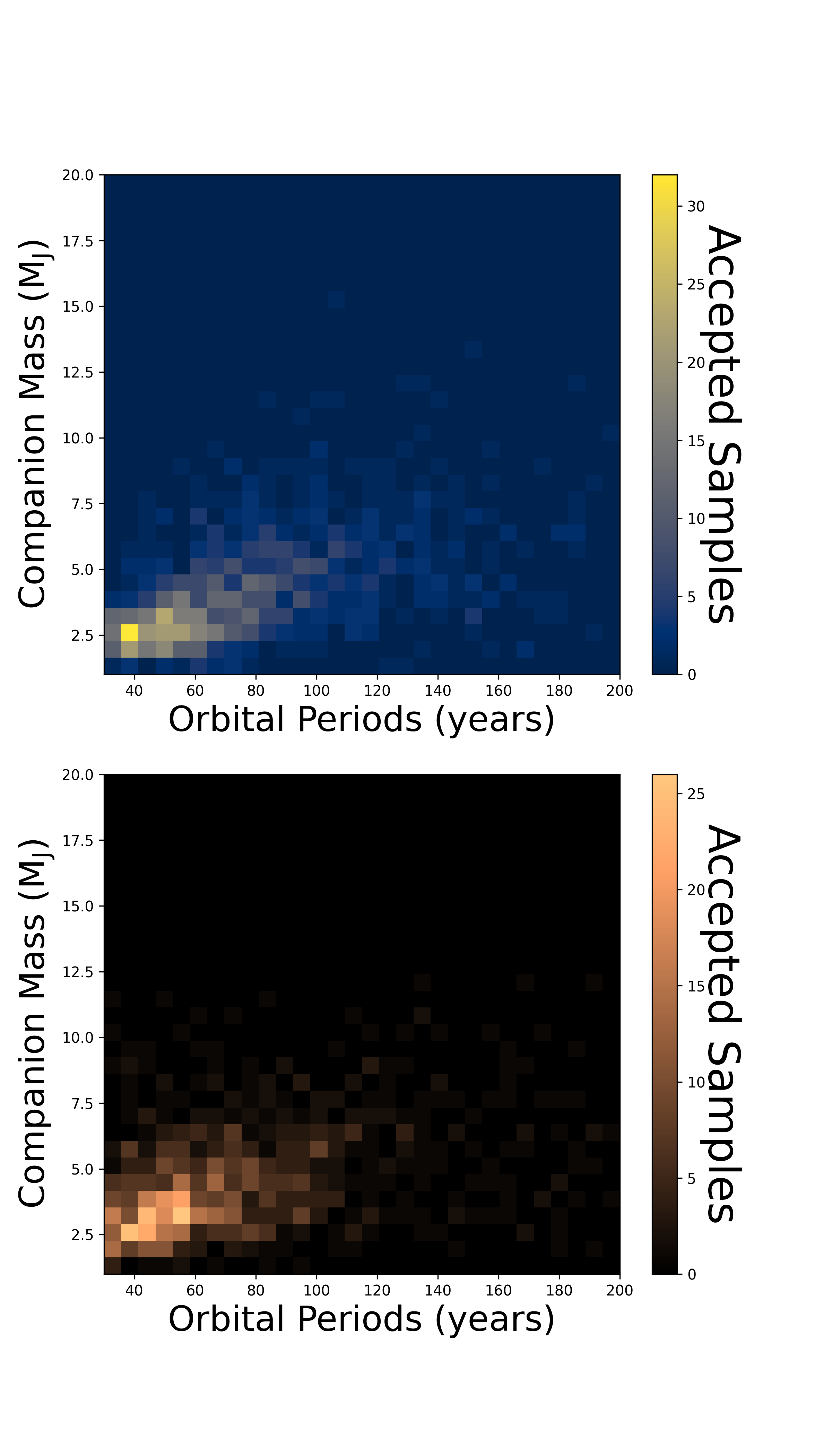}
    \caption{Top: rejection sampling for a variety of masses and orbital periods of the long period companion, where inclination is fixed at 90 degrees. Bottom: same as above, but with inclination allowed to vary. Due to the lack of sizable astrometric signatures in Gaia data, we can rule out many long period and high mass objects.}
    \label{fig:rejection_sampling}
\end{figure}

A combination of RVs and Gaia astrometry \citep{gaia23} can be used to constrain the parameters of long period companions \citep[e.g.,][]{lubin22, endl22, sozzetti23}. Since we only observe a linear drift in RVs, formal statistical analyses like MCMCs cannot reliably converge on a two-planet fit, and would produce poorly constrained posteriors. In order to make a reasonable estimate for the parameter space in which an outer companion could exist, we implement a model based on rejection sampling \citep{blunt17} that considers many randomly sampled trial orbits and accepts those that pass our acceptance criteria. We also use the same acceptance criteria, which accepts a trial if the likelihood, estimated as $0.5*\exp(-\chi^{2}/2)$ is higher than a random number between 0 and 1. We stop our sampling once 1000 trial orbits have been accepted. In addition to our residual RV slope, our rejection sampling imposes an agreement between a trial orbit's induced astrometric signal and the calibrated absolute astrometry from the Hipparcos-Gaia Catalog of Accelerations \citep[HGCA;][]{brandt18,brandt21}. The low astrometric signature of Kepler-21 rules out most high mass objects, and we can constrain mass and period modestly well. We perform fits where the companion has a fixed 90 degree inclination, and where its inclination is allowed to float. We show a plot of our rejection sampling in Figure \ref{fig:rejection_sampling}. We estimate a mass of 3.7$^ {+2.5}_{-1.3}$ M$_{\rm{J}}$ and an orbital period of 70.0$^{+52.7}_{-26.4}$ yr in the first case, and a mass of 4.0$^{+2.4}_{-1.3}$ M$_{\rm{J}}$ and period 62.7$^{+49.6}_{-21.8}$ yr in the latter case. In both cases, more than 97$\%$ of our samples are less than 10 M$_{J}$, which justifies designating the object imposing the long-term RV trend on Kepler-21 a candidate super-Jupiter planet, Kepler-21 (c). 

\begin{deluxetable*}{lccc}[h]
\label{tab:model_comparison}
\tabletypesize{\small}
\tablecaption{Model Comparisons}
\tablehead{\colhead{~~~Fit} & \colhead{Kepler-train BIC} & \colhead{TESS-train BIC} & \colhead{Joint Kepler-TESS-RV BIC}
}
\startdata
No Trend & 953.27 & 899.82 & -1193651 \\
Linear Trend & \textbf{890.06} & \textbf{878.89} & \textbf{-1193695} \\
Linear + Quadratic Trend & 906.12 & 875.95 & -1193627 \\
Two Planet Model & 906.49 & 895.98 & - \\
\enddata
\end{deluxetable*}

\section{Discussion}
\label{sec:discussion}

Our investigation and analysis of Kepler-21 had two goals: an updated, more precise description of the Kepler-21 system, and an investigation into the use of photometry to mitigate stellar variability in RVs.

\subsection{Adopted Fit}
\label{sec:adopted}

We focus on one set of posteriors when discussing the Kepler-21 planetary system. Two fits, TESS-trained RV and Joint Kepler-TESS-RV, stand out as the most successful at constraining the mass of Kepler-21 b. We choose the latter as our "adopted fit," as it is more complete than the former, utilizing more data and returning a larger set of posteriors. It also more correctly identifies the rotation period of the system. Consequently, our plots in Figure \ref{fig:RV_fit} and \ref{fig:MR} utilize this set of posterior values.

\subsection{Refining Fits for Kepler-21 b}
\label{sec:fits}

The photometry of Kepler-21 was first analyzed in \cite{howell12}, and a joint photometry-RV fit was performed in LM16. Despite a large number of RVs, stellar variability made precise recovery of the planet mass challenging. The final mass recovered in LM16 using a joint HIRES + HARPS-N RV fit was 5.08 $\pm$ 1.72 M$_{\oplus}$, which is slightly less than 3$\sigma$ significant, often considered the minimum for a ``significant" detection. One of our goals was to raise this to a higher significance threshold, and to better understand where Kepler-21 b falls in parameter space. \cite{bonomo23} recently improved the mass measurement of Kepler-21 b to $>$ 5$\sigma$ in a survey focused on Kepler systems.

Our fits utilize additional HIRES RVs and NEID data. While we cannot claim an improved mass precision, we do characterize the system with the most detail to date, and we provide significantly improved constraints on the additional companion described in \S \ref{sec:trend}. Our full fit posteriors are available in Tables \ref{tab:rv_posteriors} and \ref{tab:joint_posteriors}.

As explained in \S \ref{sec:transit}, including TESS photometry did not improve the measured orbital parameters, in particular the orbital period and time of inferior conjunction. The design of Kepler already allowed for pristine recovery of both values, with errors reported in LM16 of 0.24 seconds and 71 seconds, respectively. Our orbital period precision is identical to that in LM16, and the uncertainty in our measured transit time is actually worse. Why might this be? We first ran our exact same model, but with Kepler only, to see if adding TESS indeed improved our recovered parameters. When not utilizing TESS, our orbital period and transit time measurements are actually less precise than in LM16, despite using the same photometric dataset. We conclude that adding TESS does indeed improve precision on both posterior values, and that something about our model or parameter estimation is performing worse than in LM16. 

One explanation is our treatment of stellar variability. LM16 utilized the data validation (DV) lightcurve produced by the Kepler DV pipeline in their analysis. This lightcurve sees aggressive detrending for any periodic signals longer than the transit duration \citep{jenkins10,wu10}. The result is a pristine lightcurve, but stellar astrophysics cannot be extracted. We utilize a newer version of this reduction that does not remove long period variability, and we use a GP model to remove the residual, non-transit signals. Our motivation is to extract information about the stellar variability in the RVs from the photometry, and so we require the preservation of longer period signals. As a result, our model fits see a simultaneous GP + transit fit, which increases the uncertainty in the exact transit time.

\begin{figure}
    \centering
    \includegraphics[width=0.5\textwidth]{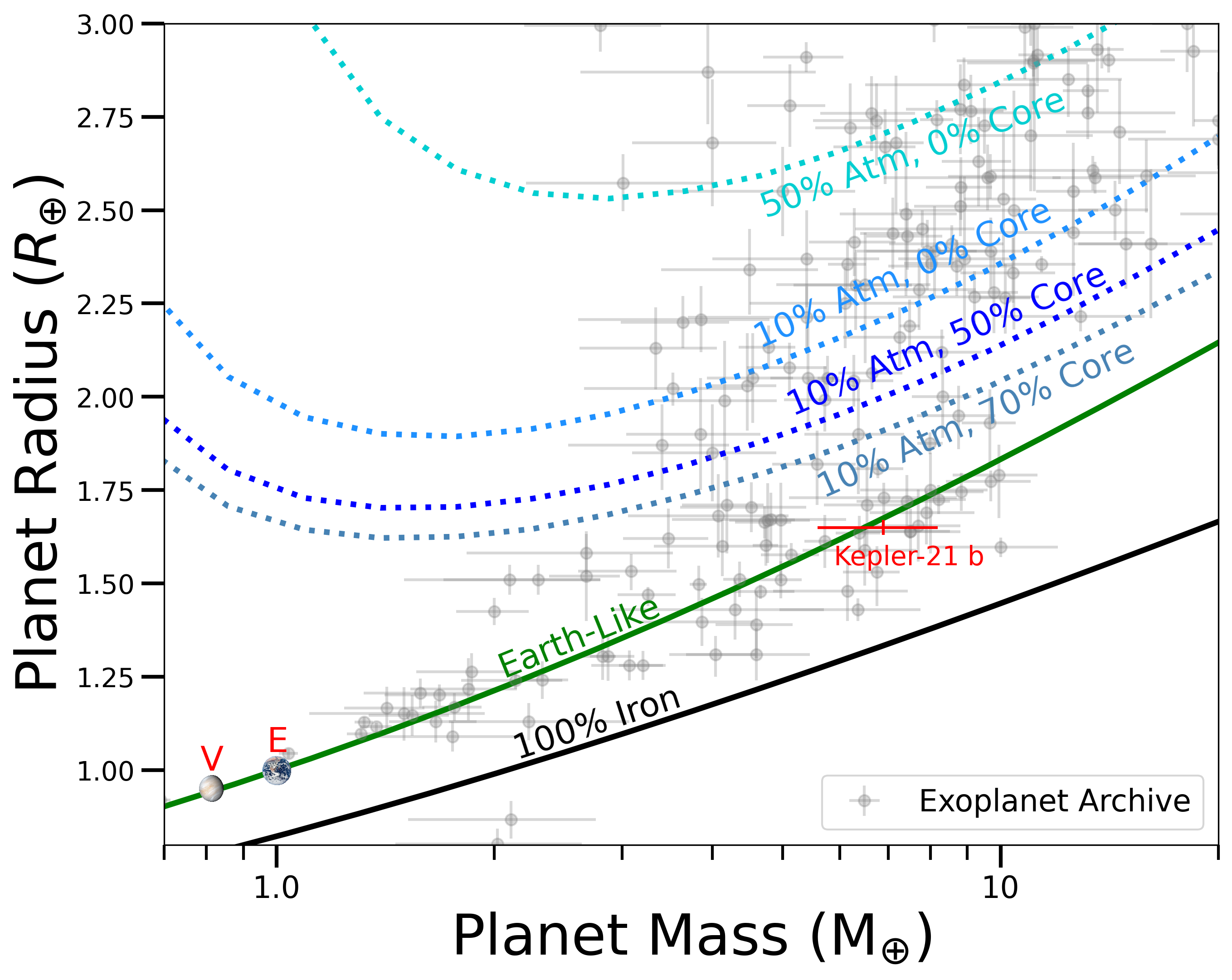}
    \caption{Correlation between mass and radius of known exoplanets with a measured mass and radius, taken from the \texttt{NASA Exoplanet Archive} on 12 August 2023. We only use planets with precisely measured masses (M$_{p}$/$\sigma$ $>$ 3). We add our new mass and radius measurements for Kepler-21 b in red. We additionally add theoretical planet compositions. Earth and iron compositions are taken from \cite{zeng19}, and we extract irradiated water compositions from \cite{aguichine21}. We use "Atm" to indicate the percentage of water in the atmosphere, and "Core" to indicate the percentage in the planet core. Kepler-21 b's placement in the radius valley made it a candidate ``Water World," though it seems such scenarios can likely be ruled out.}
    \label{fig:MR}
\end{figure}

With a newly measured planetary mass, we can rule out Kepler-21 b as a ``water world." (Figure \ref{fig:MR}). Its mass (6.68$\pm$1.4 M$_{\oplus}$) and radius (1.65$\pm$0.02 R$_{\oplus}$) place it on the upper edge of  ``water world" candidates, which \cite{luque22} denote as a regime of planet mass between 3 and 6 M$_{\oplus}$ and 1.5 - 2.0 R$_{\oplus}$. Kepler-21 b might have formed beyond the ice line and migrated inwards throughout its lifetime, retaining water and falling into this regime \citep{zeng19, luque22}. Such planets retain an atmosphere composed partially of gaseous water, and can even retain water in their core. \cite{aguichine21} model theoretical compositions for highly irradiated water worlds, and we plot selected curves in Figure \ref{fig:MR}. Kepler-21 b is too massive for it to contain any significant quantity of water in its atmosphere or in its core. Most likely, the evolution of its host star stripped the atmosphere of the planet, leaving it a bare rock. 

Atmospheric observations might provide an independent verification that the system lacks appreciable quantities of water, though Kepler-21 b is not a strong candidate for JWST. We estimate a transmission spectroscopy metric \citep[TSM;][]{kempton18} of 16.4 for Kepler-21. The TSM is a simple, first order approximation for amenability to JWST atmospheric observation. In its radius regime, \cite{kempton18} suggest at least a TSM of 92 to justify study, making Kepler-21 b far from ideal. The biggest contributor to Kepler-21 b's low TSM is the large radius of its evolved F type host star, which makes any atmospheric signal small.

\subsection{RVs Trained with Photometry}

Our second goal was to investigate the effectiveness of mitigating stellar variability in RVs using different photometric datasets. This exploration of Kepler-21 is a precursor and test case for a wider analysis of Kepler/K2/TESS targets that saw simultaneous NEID RV observations. 

A few of the questions we are interested in exploring are 1) How long are lightcurves useful for dealing with astrophysical RV noise? Are Kepler lightcurves still valuable? By extension, will TESS lightcurves be useful years from now? 2) Does the simultaneity of our NEID data mitigate the limitations of TESS baseline and precision? 3) Do joint Kepler-TESS models improve mass precision, or do the differences in sampling, age, and reduction make the noise models too different?

The broader analysis mentioned above will eventually shed more light on all of these questions, though with Kepler-21 we explore a single case deeply.

First, our analysis indicates that Kepler lightcurves are still useful. Joint fits with Kepler GP parameters sharing information with RV GP parameters typically recover the most precise mass for planet b. In the RV-only fits, training on Kepler photometry is noticeably worse than training on TESS, though as we discuss in \S \ref{sec:training}, this is probably due to the Kepler training adhering to the wrong rotation period. As seen in Figures \ref{fig:IR_fixed} and \ref{fig:IR_vary}, Kepler training helps recover additional planets broadly across RV amplitude and orbital period space.

Interestingly, injected planets with longer periods seem to benefit more from training on photometry, especially in the injected transiting planet case. This is probably best explained by the sparsity of RV datasets: longer period injected planets will see their orbital periods less-well resolved via RV observations. Since an untrained GP model is generally more flexible than our trained GP models, the untrained case will likely often confuse the sparse injected planet with a stellar activity signal.

Despite being over ten years old, Kepler photometry is highly precise, and huge in quantity. This trumps the more recent TESS data, at least for this system. This may not be true for every system, however. We did not achieve as many simultaneous NEID RVs as would have been ideal, and unlike many Kepler systems, Kepler-21 has precise RVs simultaneous with Kepler. Due to the higher rotational velocity of Kepler-21, NEID is not seeing higher precision than older RVs, such as HARPS-N or HIRES. We can also conclude that TESS photometry may still be useful ten or more years in the future, though TESS's specific usefulness will likely depend on the existence of other photometric datasets \citep[e.g. PLATO;][]{rauer14} at the time.

Second, does the simultaneity of NEID with TESS mitigate some of the downsides of TESS photometry? PDCSAP photometry removes longer period signals, especially those on the order of half a 27-day TESS cycle. With a rotation period of 12.6 days, Kepler-21 is near the upper limit of what might be possible to find in TESS PDCSAP. This obviously makes TESS less useful for constraining stellar variability in RVs, since longer period signals in RVs still remain. However, unlike older Kepler photometry, simultaneous TESS photometry can give us information about spot complexes as the RV data are taken, which might be advantageous. Indeed the RV fits trained on TESS photometry were typically more precise than those trained on Kepler, though this may be the result of the erroneous period detection mentioned in \S \ref{sec:training}. Nonetheless, we can say that training on TESS certainly can be effective, though training on both jointly has its advantages.

Third, should we use a joint Kepler-TESS model when informing RVs? We examine our model that performs such a fit, and how it compares to alternatives. In particular, we ask the question: should we train an activity model on photometry, and then perform an RV-only fit? Or is there an advantage to performing these steps simultaneously? We use the recovered RV semi-amplitude as our primary comparison point, because we are most interested in which method best removes stellar activity contamination in the RVs. This activity contamination has the most pressing effect on the recovery of this observable.

Comparing the most precise mass measurements in each case, our TESS-trained RV fit has a semi-amplitude of 2.59$\pm$0.46 m s$^{-1}$, while joint Kepler-TESS-RV fits have an amplitude of 2.49$\pm$0.47 m s$^{-1}$. The former fit has a strictly higher precision, though the two results are extremely close. The methods recover the RV amplitude indistinguishably well. However, joint fits are generally more computationally costly, and so this may be seen as an endorsement of RV fits trained on photometry. On the other hand, the joint fits recover the true rotation period of the system, and this may be seen as an endorsement of that method.

\subsection{Kepler-21 (c)?}

If it is a planet, Kepler-21 (c) would have the longest known orbital period of any planet with a known transiting planet companion. We estimate the orbital separation of Kepler-21 (c) in the more general non-fixed-inclination case as 17.7$\pm$1.6 AU. Kepler-21 is  108.5$\pm$0.4 pc \citep{bailer-jones18,gaia23} from Earth, giving Kepler-21 (c) an angular separation of 160 mas. Other giant planets have been imaged at this separation \citep[PDS 70 b; 175.8 mas][]{wang20}, though the age of Kepler-21 makes that prospect  more challenging. Unlike most planets imaged today, Kepler-21 (c) has likely lost most of its heat of formation, and would likely need to be imaged using reflected light from its host star. We calculate the contrast Kepler-21 (c) would likely exhibit with respect to its host star using equation \ref{eqn:contrast} \citep{li21}.

\begin{equation}
    \label{eqn:contrast}
    \centering
    \epsilon = A * \frac{1}{\pi} * \frac{R_{p}^2}{a^{2}}
\end{equation}

Above, $A$ is the albedo of the planet, $a$ its semi-major axis, and R$_{p}$ its radius. We use the mass-radius relationship in \cite{chen17} to estimate a planet radius for Kepler-21 (c) of 1.16$^{+0.24}_{-0.20}$ R$_{\rm{J}}$. Using the calculated semi-major axis of 17.7$\pm$1.6 AU, we estimate a contrast of 3.1e-11$\pm$1.6e-11 using an albedo of 0.1. In the highly reflective case of albedo=1, this only increases the contrast by one order of magnitude, 3.1e-10$\pm$1.6e-10. This is below the expected atmospheric contrast limit for even extremely large telescopes \citep[1e-8;][]{stapelfeldt06}, and Kepler-21 (c) will not likely ever be imaged from the ground, though the proposed Habitable Worlds Observatory (HWO) has a targeted contrast limit of 1e-10 \citep{decadal20,harada24}, making the high-albedo case potentially observable. 

While both transmission spectroscopy of the inner planet, and direct imaging spectroscopy of the outer candidate are at the edge of even future detection limits, it is possible that both observations might be taken someday. An intra-system comparison of the atmospheric compositions of the two bodies could reveal a great deal about the formation and evolution of the system, and would be highly valuable.

\section{Summary}
\label{sec:summary}

We revisit the Kepler-21 system and perform an in-depth analysis of the best methods to mitigate stellar activity with photometry. We compare the results of RV fits trained on Kepler and TESS data, as well as a variety of joint fits. We also perform an injection recovery test to determine if Kepler or TESS photometry is better at disentangling injected planet signals from stellar activity. Our results show that training activity models on Kepler benefits in the discovery of new planets, though our TESS-trained models recover a more precise mass for Kepler-21 b. We further confirm the nature of Kepler-21 b as a rocky, terrestrial planet in the radius valley, and we strongly identify a long-period companion in the system.

\section{Acknowledgements}

This paper includes data collected by the TESS mission. Funding for the TESS mission is provided by the NASA's Science Mission Directorate.

This paper includes data collected by the Kepler mission and obtained from the MAST data archive at the Space Telescope Science Institute (STScI). Funding for the Kepler mission is provided by the NASA Science Mission Directorate. STScI is operated by the Association of Universities for Research in Astronomy, Inc., under NASA contract NAS 5–26555.


Some of the data presented herein were obtained at Keck Observatory, which is a private 501(c)3 non-profit organization operated as a scientific partnership among the California Institute of Technology, the University of California, and the National Aeronautics and Space Administration. The Observatory was made possible by the generous financial support of the W. M. Keck Foundation.

The authors wish to recognize and acknowledge the very significant cultural role and reverence that the summit of Maunakea has always had within the Native Hawaiian community. We are most fortunate to have the opportunity to conduct observations from this mountain.


Based on observations at Kitt Peak National Observatory at NSF’s NOIRLab (NOIRLab Prop. ID 2021A-0396; PI: C. Beard; NOIRLab Prop. ID 2022A-377087; PI: C. Beard; NOIRLab Prop. ID 2022A-399413; PI: C. Beard;), which is managed by the Association of Universities for Research in Astronomy (AURA) under a cooperative agreement with the National Science Foundation. The authors are honored to be permitted to conduct astronomical research on Iolkam Du’ag (Kitt Peak), a mountain with particular significance to the Tohono O’odham.

This work was partially supported by NASA grant 80NSSC22K0120 to support Guest Investigator programs for TESS Cycle 4. 

This work was partially support by the Future Investigators in NASA Earth and Space Science and Technology (FINESST) program Grant No. 80NSSC22K1754.

This research made use of \textsf{exoplanet} \citep{exoplanet:joss, exoplanet:zenodo} and its dependencies \citep{exoplanet:foremanmackey17, exoplanet:foremanmackey18, exoplanet:agol20, exoplanet:arviz, exoplanet:astropy13, exoplanet:astropy18, exoplanet:luger18, exoplanet:pymc3, exoplanet:theano}.


This work utilized the infrastructure for high-performance and high-throughput computing, research data storage and analysis, and scientific software tool integration built, operated, and updated by the Research Cyberinfrastructure Center (RCIC) at the University of California, Irvine (UCI). The RCIC provides cluster-based systems, application software, and scalable storage to directly support the UCI research community. https://rcic.uci.edu

J.M.A.M. acknowledges support from the National Science Foundation Graduate Research Fellowship Program under Grant No. DGE-1842400.

The Center for Exoplanets and Habitable Worlds and the Penn State Extraterrestrial Intelligence Center are supported by the Pennsylvania State University and the Eberly College of Science.  

This research was carried out, in part, at the Jet Propulsion Laboratory and the California Institute of Technology under a contract with the National Aeronautics and Space Administration and funded through the President’s and Director’s Research \& Development Fund Program.
\bibliography{bibliography}

\begin{thebibliography}{}
\expandafter\ifx\csname natexlab\endcsname\relax\def\natexlab#1{#1}\fi
\providecommand{\url}[1]{\href{#1}{#1}}

\bibitem[{{Agol} {et~al.}(2020){Agol}, {Luger}, \&
  {Foreman-Mackey}}]{exoplanet:agol20}
{Agol}, E., {Luger}, R., \& {Foreman-Mackey}, D. 2020, \aj, 159, 123

\bibitem[{{Aguichine} {et~al.}(2021){Aguichine}, {Mousis}, {Deleuil}, \&
  {Marcq}}]{aguichine21}
{Aguichine}, A., {Mousis}, O., {Deleuil}, M., \& {Marcq}, E. 2021, \apj, 914,
  84

\bibitem[{{Aigrain} {et~al.}(2012){Aigrain}, {Pont}, \& {Zucker}}]{aigrain12}
{Aigrain}, S., {Pont}, F., \& {Zucker}, S. 2012, \mnras, 419, 3147

\bibitem[{{Akana Murphy} {et~al.}(2023){Akana Murphy}, {Batalha}, {Scarsdale},
  {Isaacson}, {Ciardi}, {Gonzales}, {Giacalone}, {Twicken}, {Dattilo},
  {Fetherolf}, {Rubenzahl}, {Crossfield}, {Dressing}, {Fulton}, {Howard},
  {Huber}, {Kane}, {Petigura}, {Robertson}, {Roy}, {Weiss}, {Beard}, {Chontos},
  {Dai}, {Rice}, {Van Zandt}, {Lubin}, {Blunt}, {Polanski}, {Behmard}, {Dalba},
  {Hill}, {Rosenthal}, {Brinkman}, {Mayo}, {Turtelboom}, {Angelo},
  {Mo{\v{c}}nik}, {MacDougall}, {Pidhorodetska}, {Tyler}, {Kosiarek},
  {Holcomb}, {Louden}, {Hirsch}, {Gilbert}, {Anderson}, \&
  {Valenti}}]{murphy23}
{Akana Murphy}, J.~M., {Batalha}, N.~M., {Scarsdale}, N., {et~al.} 2023, \aj,
  166, 153

\bibitem[{{Ambikasaran} {et~al.}(2015){Ambikasaran}, {Foreman-Mackey},
  {Greengard}, {Hogg}, \& {O'Neil}}]{ambikasaran15}
{Ambikasaran}, S., {Foreman-Mackey}, D., {Greengard}, L., {Hogg}, D.~W., \&
  {O'Neil}, M. 2015, IEEE Transactions on Pattern Analysis and Machine
  Intelligence, 38, 252

\bibitem[{{Astropy Collaboration} {et~al.}(2013){Astropy Collaboration},
  {Robitaille}, {Tollerud}, {Greenfield}, {Droettboom}, {Bray}, {Aldcroft},
  {Davis}, {Ginsburg}, {Price-Whelan}, {Kerzendorf}, {Conley}, {Crighton},
  {Barbary}, {Muna}, {Ferguson}, {Grollier}, {Parikh}, {Nair}, {Unther},
  {Deil}, {Woillez}, {Conseil}, {Kramer}, {Turner}, {Singer}, {Fox}, {Weaver},
  {Zabalza}, {Edwards}, {Azalee Bostroem}, {Burke}, {Casey}, {Crawford},
  {Dencheva}, {Ely}, {Jenness}, {Labrie}, {Lim}, {Pierfederici}, {Pontzen},
  {Ptak}, {Refsdal}, {Servillat}, \& {Streicher}}]{exoplanet:astropy13}
{Astropy Collaboration}, {Robitaille}, T.~P., {Tollerud}, E.~J., {et~al.} 2013,
  \aap, 558, A33

\bibitem[{{Astropy Collaboration} {et~al.}(2018){Astropy Collaboration},
  {Price-Whelan}, {Sip{\H o}cz}, {G{\"u}nther}, {Lim}, {Crawford}, {Conseil},
  {Shupe}, {Craig}, {Dencheva}, {Ginsburg}, {VanderPlas}, {Bradley},
  {P{\'e}rez-Su{\'a}rez}, {de Val-Borro}, {Aldcroft}, {Cruz}, {Robitaille},
  {Tollerud}, {Ardelean}, {Babej}, {Bach}, {Bachetti}, {Bakanov}, {Bamford},
  {Barentsen}, {Barmby}, {Baumbach}, {Berry}, {Biscani}, {Boquien}, {Bostroem},
  {Bouma}, {Brammer}, {Bray}, {Breytenbach}, {Buddelmeijer}, {Burke},
  {Calderone}, {Cano Rodr{\'{\i}}guez}, {Cara}, {Cardoso}, {Cheedella},
  {Copin}, {Corrales}, {Crichton}, {D'Avella}, {Deil}, {Depagne}, {Dietrich},
  {Donath}, {Droettboom}, {Earl}, {Erben}, {Fabbro}, {Ferreira}, {Finethy},
  {Fox}, {Garrison}, {Gibbons}, {Goldstein}, {Gommers}, {Greco}, {Greenfield},
  {Groener}, {Grollier}, {Hagen}, {Hirst}, {Homeier}, {Horton}, {Hosseinzadeh},
  {Hu}, {Hunkeler}, {Ivezi{\'c}}, {Jain}, {Jenness}, {Kanarek}, {Kendrew},
  {Kern}, {Kerzendorf}, {Khvalko}, {King}, {Kirkby}, {Kulkarni}, {Kumar},
  {Lee}, {Lenz}, {Littlefair}, {Ma}, {Macleod}, {Mastropietro}, {McCully},
  {Montagnac}, {Morris}, {Mueller}, {Mumford}, {Muna}, {Murphy}, {Nelson},
  {Nguyen}, {Ninan}, {N{\"o}the}, {Ogaz}, {Oh}, {Parejko}, {Parley}, {Pascual},
  {Patil}, {Patil}, {Plunkett}, {Prochaska}, {Rastogi}, {Reddy Janga},
  {Sabater}, {Sakurikar}, {Seifert}, {Sherbert}, {Sherwood-Taylor}, {Shih},
  {Sick}, {Silbiger}, {Singanamalla}, {Singer}, {Sladen}, {Sooley},
  {Sornarajah}, {Streicher}, {Teuben}, {Thomas}, {Tremblay}, {Turner},
  {Terr{\'o}n}, {van Kerkwijk}, {de la Vega}, {Watkins}, {Weaver}, {Whitmore},
  {Woillez}, {Zabalza}, \& {Astropy Contributors}}]{exoplanet:astropy18}
{Astropy Collaboration}, {Price-Whelan}, A.~M., {Sip{\H o}cz}, B.~M., {et~al.}
  2018, \aj, 156, 123

\bibitem[{{Bailer-Jones} {et~al.}(2018){Bailer-Jones}, {Rybizki}, {Fouesneau},
  {Mantelet}, \& {Andrae}}]{bailer-jones18}
{Bailer-Jones}, C.~A.~L., {Rybizki}, J., {Fouesneau}, M., {Mantelet}, G., \&
  {Andrae}, R. 2018, \aj, 156, 58

\bibitem[{{Bastien} {et~al.}(2014){Bastien}, {Stassun}, {Pepper}, {Wright},
  {Aigrain}, {Basri}, {Johnson}, {Howard}, \& {Walkowicz}}]{bastien14}
{Bastien}, F.~A., {Stassun}, K.~G., {Pepper}, J., {et~al.} 2014, \aj, 147, 29

\bibitem[{{Beard} {et~al.}(2022){Beard}, {Robertson}, {Kanodia},
  {Libby-Roberts}, {Ca{\~n}as}, {Gupta}, {Holcomb}, {Jones}, {Kobulnicky},
  {Lin}, {Lubin}, {Maney}, {Parker}, {Stef{\'a}nsson}, {Cochran}, {Endl},
  {Hebb}, {Mahadevan}, {Wisniewski}, {Bender}, {Diddams}, {Everett},
  {Fredrick}, {Halverson}, {Hearty}, {Metcalf}, {Monson}, {Ninan}, {Roy},
  {Schutte}, {Schwab}, \& {Terrien}}]{beard22}
{Beard}, C., {Robertson}, P., {Kanodia}, S., {et~al.} 2022, \aj, 163, 286

\bibitem[{{Beard} {et~al.}(2024){Beard}, {Robertson}, {Dai}, {Holcomb},
  {Lubin}, {Akana Murphy}, {Batalha}, {Blunt}, {Crossfield}, {Dressing},
  {Fulton}, {Howard}, {Huber}, {Isaacson}, {Kane}, {Nowak}, {Petigura}, {Roy},
  {Rubenzahl}, {Weiss}, {Barrena}, {Behmard}, {Brinkman}, {Carleo}, {Chontos},
  {Dalba}, {Fetherolf}, {Giacalone}, {Hill}, {Kawauchi}, {Korth}, {Luque},
  {MacDougall}, {Mayo}, {Mo{\v{c}}nik}, {Morello}, {Murgas}, {Orell-Miquel},
  {Palle}, {Polanski}, {Rice}, {Scarsdale}, {Tyler}, \& {Van Zandt}}]{beard24}
{Beard}, C., {Robertson}, P., {Dai}, F., {et~al.} 2024, \aj, 167, 70

\bibitem[{{Blunt} {et~al.}(2017){Blunt}, {Nielsen}, {De Rosa}, {Konopacky},
  {Ryan}, {Wang}, {Pueyo}, {Rameau}, {Marois}, {Marchis}, {Macintosh},
  {Graham}, {Duch{\^e}ne}, \& {Schneider}}]{blunt17}
{Blunt}, S., {Nielsen}, E.~L., {De Rosa}, R.~J., {et~al.} 2017, \aj, 153, 229

\bibitem[{{Blunt} {et~al.}(2023){Blunt}, {Carvalho}, {David}, {Beichman},
  {Zink}, {Gaidos}, {Behmard}, {Bouma}, {Cody}, {Dai}, {Foreman-Mackey},
  {Grunblatt}, {Howard}, {Kosiarek}, {Knutson}, {Rubenzahl}, {Beard},
  {Chontos}, {Giacalone}, {Hirano}, {Johnson}, {Lubin}, {Akana Murphy},
  {Petigura}, {Van Zandt}, \& {Weiss}}]{blunt23}
{Blunt}, S., {Carvalho}, A., {David}, T.~J., {et~al.} 2023, arXiv e-prints,
  arXiv:2306.08145

\bibitem[{{Bonomo} {et~al.}(2023){Bonomo}, {Dumusque}, {Massa}, {Mortier},
  {Bongiolatti}, {Malavolta}, {Sozzetti}, {Buchhave}, {Damasso}, {Haywood},
  {Morbidelli}, {Latham}, {Molinari}, {Pepe}, {Poretti}, {Udry}, {Affer},
  {Boschin}, {Charbonneau}, {Cosentino}, {Cretignier}, {Ghedina}, {Lega},
  {L{\'o}pez-Morales}, {Margini}, {Mart{\'\i}nez Fiorenzano}, {Mayor},
  {Micela}, {Pedani}, {Pinamonti}, {Rice}, {Sasselov}, {Tronsgaard}, \&
  {Vanderburg}}]{bonomo23}
{Bonomo}, A.~S., {Dumusque}, X., {Massa}, A., {et~al.} 2023, \aap, 677, A33

\bibitem[{{Borucki} {et~al.}(2010){Borucki}, {Koch}, {Basri}, {Batalha},
  {Brown}, {Caldwell}, {Caldwell}, {Christensen-Dalsgaard}, {Cochran},
  {DeVore}, {Dunham}, {Dupree}, {Gautier}, {Geary}, {Gilliland}, {Gould},
  {Howell}, {Jenkins}, {Kondo}, {Latham}, {Marcy}, {Meibom}, {Kjeldsen},
  {Lissauer}, {Monet}, {Morrison}, {Sasselov}, {Tarter}, {Boss}, {Brownlee},
  {Owen}, {Buzasi}, {Charbonneau}, {Doyle}, {Fortney}, {Ford}, {Holman},
  {Seager}, {Steffen}, {Welsh}, {Rowe}, {Anderson}, {Buchhave}, {Ciardi},
  {Walkowicz}, {Sherry}, {Horch}, {Isaacson}, {Everett}, {Fischer}, {Torres},
  {Johnson}, {Endl}, {MacQueen}, {Bryson}, {Dotson}, {Haas}, {Kolodziejczak},
  {Van Cleve}, {Chandrasekaran}, {Twicken}, {Quintana}, {Clarke}, {Allen},
  {Li}, {Wu}, {Tenenbaum}, {Verner}, {Bruhweiler}, {Barnes}, \&
  {Prsa}}]{borucki10}
{Borucki}, W.~J., {Koch}, D., {Basri}, G., {et~al.} 2010, Science, 327, 977

\bibitem[{{Brandt}(2018)}]{brandt18}
{Brandt}, T.~D. 2018, \apjs, 239, 31

\bibitem[{{Brandt}(2021)}]{brandt21}
---. 2021, \apjs, 254, 42

\bibitem[{{Butler} \& {Marcy}(1996)}]{butler96b}
{Butler}, R.~P., \& {Marcy}, G.~W. 1996, \apjl, 464, L153

\bibitem[{{Butler} {et~al.}(1997){Butler}, {Marcy}, {Williams}, {Hauser}, \&
  {Shirts}}]{butler97}
{Butler}, R.~P., {Marcy}, G.~W., {Williams}, E., {Hauser}, H., \& {Shirts}, P.
  1997, \apjl, 474, L115

\bibitem[{{Butler} {et~al.}(1996){Butler}, {Marcy}, {Williams}, {McCarthy},
  {Dosanjh}, \& {Vogt}}]{butler96a}
{Butler}, R.~P., {Marcy}, G.~W., {Williams}, E., {et~al.} 1996, \pasp, 108, 500

\bibitem[{{Butler} {et~al.}(2017){Butler}, {Vogt}, {Laughlin}, {Burt},
  {Rivera}, {Tuomi}, {Teske}, {Arriagada}, {Diaz}, {Holden}, \&
  {Keiser}}]{butler17}
{Butler}, R.~P., {Vogt}, S.~S., {Laughlin}, G., {et~al.} 2017, \aj, 153, 208

\bibitem[{{Cale} {et~al.}(2021){Cale}, {Reefe}, {Plavchan}, {Tanner}, {Gaidos},
  {Gagn{\'e}}, {Gao}, {Kane}, {B{\'e}jar}, {Lodieu}, {Anglada-Escud{\'e}},
  {Ribas}, {Pall{\'e}}, {Quirrenbach}, {Amado}, {Reiners}, {Caballero}, {Rosa
  Zapatero Osorio}, {Dreizler}, {Howard}, {Fulton}, {Xuesong Wang}, {Collins},
  {El Mufti}, {Wittrock}, {Gilbert}, {Barclay}, {Klein}, {Martioli},
  {Wittenmyer}, {Wright}, {Addison}, {Hirano}, {Tamura}, {Kotani}, {Narita},
  {Vermilion}, {Lee}, {Geneser}, {Teske}, {Quinn}, {Latham}, {Esquerdo},
  {Calkins}, {Berlind}, {Zohrabi}, {Stibbards}, {Kotnana}, {Jenkins},
  {Twicken}, {Henze}, {Kidwell}, {Burke}, {Villase{\~n}or}, \& {Boyd}}]{cale21}
{Cale}, B.~L., {Reefe}, M., {Plavchan}, P., {et~al.} 2021, \aj, 162, 295

\bibitem[{{Cegla} {et~al.}(2019){Cegla}, {Watson}, {Shelyag}, {Mathioudakis},
  \& {Moutari}}]{cegla19}
{Cegla}, H.~M., {Watson}, C.~A., {Shelyag}, S., {Mathioudakis}, M., \&
  {Moutari}, S. 2019, \apj, 879, 55

\bibitem[{{Chaplin} {et~al.}(2019){Chaplin}, {Cegla}, {Watson}, {Davies}, \&
  {Ball}}]{chaplin19}
{Chaplin}, W.~J., {Cegla}, H.~M., {Watson}, C.~A., {Davies}, G.~R., \& {Ball},
  W.~H. 2019, \aj, 157, 163

\bibitem[{{Chen} \& {Kipping}(2017)}]{chen17}
{Chen}, J., \& {Kipping}, D. 2017, \apj, 834, 17

\bibitem[{{Claret} {et~al.}(2012){Claret}, {Hauschildt}, \& {Witte}}]{claret12}
{Claret}, A., {Hauschildt}, P.~H., \& {Witte}, S. 2012, \aap, 546, A14

\bibitem[{{Claret} {et~al.}(2013){Claret}, {Hauschildt}, \& {Witte}}]{claret13}
---. 2013, \aap, 552, A16

\bibitem[{{Cosentino} {et~al.}(2012){Cosentino}, {Lovis}, {Pepe}, {Collier
  Cameron}, {Latham}, {Molinari}, {Udry}, {Bezawada}, {Black}, {Born},
  {Buchschacher}, {Charbonneau}, {Figueira}, {Fleury}, {Galli}, {Gallie},
  {Gao}, {Ghedina}, {Gonzalez}, {Gonzalez}, {Guerra}, {Henry}, {Horne},
  {Hughes}, {Kelly}, {Lodi}, {Lunney}, {Maire}, {Mayor}, {Micela}, {Ordway},
  {Peacock}, {Phillips}, {Piotto}, {Pollacco}, {Queloz}, {Rice}, {Riverol},
  {Riverol}, {San Juan}, {Sasselov}, {Segransan}, {Sozzetti}, {Sosnowska},
  {Stobie}, {Szentgyorgyi}, {Vick}, \& {Weber}}]{cosentino12}
{Cosentino}, R., {Lovis}, C., {Pepe}, F., {et~al.} 2012, in Society of
  Photo-Optical Instrumentation Engineers (SPIE) Conference Series, Vol. 8446,
  Ground-based and Airborne Instrumentation for Astronomy IV, ed. I.~S.
  {McLean}, S.~K. {Ramsay}, \& H.~{Takami}, 84461V

\bibitem[{{Crockett} {et~al.}(2012){Crockett}, {Mahmud}, {Prato},
  {Johns-Krull}, {Jaffe}, {Hartigan}, \& {Beichman}}]{crockett12}
{Crockett}, C.~J., {Mahmud}, N.~I., {Prato}, L., {et~al.} 2012, \apj, 761, 164

\bibitem[{Dierckx(1995)}]{dierckx95}
Dierckx, P. 1995, Curve and surface fitting with splines (Oxford University
  Press)

\bibitem[{{Dumusque} {et~al.}(2014){Dumusque}, {Boisse}, \&
  {Santos}}]{dumusque14}
{Dumusque}, X., {Boisse}, I., \& {Santos}, N.~C. 2014, \apj, 796, 132

\bibitem[{{Endl} {et~al.}(2022){Endl}, {Robertson}, {Cochran}, {MacQueen},
  {Bowler}, {Franson}, {Holcomb}, {Beard}, {Isaacson}, {Howard}, \&
  {Lubin}}]{endl22}
{Endl}, M., {Robertson}, P., {Cochran}, W.~D., {et~al.} 2022, \aj, 164, 238

\bibitem[{{Ford}(2006)}]{ford06}
{Ford}, E.~B. 2006, \apj, 642, 505

\bibitem[{{Foreman-Mackey}(2018)}]{exoplanet:foremanmackey18}
{Foreman-Mackey}, D. 2018, Research Notes of the American Astronomical Society,
  2, 31

\bibitem[{{Foreman-Mackey} {et~al.}(2017){Foreman-Mackey}, {Agol},
  {Ambikasaran}, \& {Angus}}]{exoplanet:foremanmackey17}
{Foreman-Mackey}, D., {Agol}, E., {Ambikasaran}, S., \& {Angus}, R. 2017, \aj,
  154, 220

\bibitem[{Foreman-Mackey {et~al.}(2021)Foreman-Mackey, Savel, Luger, Agol,
  Czekala, Price-Whelan, Hedges, Gilbert, Bouma, Brandt, \&
  Barclay}]{exoplanet:zenodo}
Foreman-Mackey, D., Savel, A., Luger, R., {et~al.} 2021,
  exoplanet-dev/exoplanet v0.5.1, , , doi:10.5281/zenodo.1998447.
\newblock \url{https://doi.org/10.5281/zenodo.1998447}

\bibitem[{{Foreman-Mackey} {et~al.}(2021){Foreman-Mackey}, {Luger}, {Agol},
  {Barclay}, {Bouma}, {Brandt}, {Czekala}, {David}, {Dong}, {Gilbert},
  {Gordon}, {Hedges}, {Hey}, {Morris}, {Price-Whelan}, \&
  {Savel}}]{exoplanet:joss}
{Foreman-Mackey}, D., {Luger}, R., {Agol}, E., {et~al.} 2021, arXiv e-prints,
  arXiv:2105.01994

\bibitem[{{Fulton} {et~al.}(2018){Fulton}, {Petigura}, {Blunt}, \&
  {Sinukoff}}]{fulton18}
{Fulton}, B.~J., {Petigura}, E.~A., {Blunt}, S., \& {Sinukoff}, E. 2018, \pasp,
  130, 044504

\bibitem[{{Fulton} {et~al.}(2017){Fulton}, {Petigura}, {Howard}, {Isaacson},
  {Marcy}, {Cargile}, {Hebb}, {Weiss}, {Johnson}, {Morton}, {Sinukoff},
  {Crossfield}, \& {Hirsch}}]{fulton17}
{Fulton}, B.~J., {Petigura}, E.~A., {Howard}, A.~W., {et~al.} 2017, \aj, 154,
  109

\bibitem[{{Gaia Collaboration} {et~al.}(2023){Gaia Collaboration}, {Vallenari},
  {Brown}, {Prusti}, {de Bruijne}, {Arenou}, {Babusiaux}, {Biermann},
  {Creevey}, {Ducourant}, {Evans}, {Eyer}, {Guerra}, {Hutton}, {Jordi},
  {Klioner}, {Lammers}, {Lindegren}, {Luri}, {Mignard}, {Panem}, {Pourbaix},
  {Randich}, {Sartoretti}, {Soubiran}, {Tanga}, {Walton}, {Bailer-Jones},
  {Bastian}, {Drimmel}, {Jansen}, {Katz}, {Lattanzi}, {van Leeuwen}, {Bakker},
  {Cacciari}, {Casta{\~n}eda}, {De Angeli}, {Fabricius}, {Fouesneau},
  {Fr{\'e}mat}, {Galluccio}, {Guerrier}, {Heiter}, {Masana}, {Messineo},
  {Mowlavi}, {Nicolas}, {Nienartowicz}, {Pailler}, {Panuzzo}, {Riclet}, {Roux},
  {Seabroke}, {Sordo}, {Th{\'e}venin}, {Gracia-Abril}, {Portell}, {Teyssier},
  {Altmann}, {Andrae}, {Audard}, {Bellas-Velidis}, {Benson}, {Berthier},
  {Blomme}, {Burgess}, {Busonero}, {Busso}, {C{\'a}novas}, {Carry}, {Cellino},
  {Cheek}, {Clementini}, {Damerdji}, {Davidson}, {de Teodoro}, {Nu{\~n}ez
  Campos}, {Delchambre}, {Dell'Oro}, {Esquej}, {Fern{\'a}ndez-Hern{\'a}ndez},
  {Fraile}, {Garabato}, {Garc{\'\i}a-Lario}, {Gosset}, {Haigron}, {Halbwachs},
  {Hambly}, {Harrison}, {Hern{\'a}ndez}, {Hestroffer}, {Hodgkin}, {Holl},
  {Jan{\ss}en}, {Jevardat de Fombelle}, {Jordan}, {Krone-Martins}, {Lanzafame},
  {L{\"o}ffler}, {Marchal}, {Marrese}, {Moitinho}, {Muinonen}, {Osborne},
  {Pancino}, {Pauwels}, {Recio-Blanco}, {Reyl{\'e}}, {Riello}, {Rimoldini},
  {Roegiers}, {Rybizki}, {Sarro}, {Siopis}, {Smith}, {Sozzetti}, {Utrilla},
  {van Leeuwen}, {Abbas}, {{\'A}brah{\'a}m}, {Abreu Aramburu}, {Aerts},
  {Aguado}, {Ajaj}, {Aldea-Montero}, {Altavilla}, {{\'A}lvarez}, {Alves},
  {Anders}, {Anderson}, {Anglada Varela}, {Antoja}, {Baines}, {Baker},
  {Balaguer-N{\'u}{\~n}ez}, {Balbinot}, {Balog}, {Barache}, {Barbato},
  {Barros}, {Barstow}, {Bartolom{\'e}}, {Bassilana}, {Bauchet}, {Becciani},
  {Bellazzini}, {Berihuete}, {Bernet}, {Bertone}, {Bianchi}, {Binnenfeld},
  {Blanco-Cuaresma}, {Blazere}, {Boch}, {Bombrun}, {Bossini}, {Bouquillon},
  {Bragaglia}, {Bramante}, {Breedt}, {Bressan}, {Brouillet}, {Brugaletta},
  {Bucciarelli}, {Burlacu}, {Butkevich}, {Buzzi}, {Caffau}, {Cancelliere},
  {Cantat-Gaudin}, {Carballo}, {Carlucci}, {Carnerero}, {Carrasco},
  {Casamiquela}, {Castellani}, {Castro-Ginard}, {Chaoul}, {Charlot}, {Chemin},
  {Chiaramida}, {Chiavassa}, {Chornay}, {Comoretto}, {Contursi}, {Cooper},
  {Cornez}, {Cowell}, {Crifo}, {Cropper}, {Crosta}, {Crowley}, {Dafonte},
  {Dapergolas}, {David}, {David}, {de Laverny}, {De Luise}, {De March}, {De
  Ridder}, {de Souza}, {de Torres}, {del Peloso}, {del Pozo}, {Delbo},
  {Delgado}, {Delisle}, {Demouchy}, {Dharmawardena}, {Di Matteo}, {Diakite},
  {Diener}, {Distefano}, {Dolding}, {Edvardsson}, {Enke}, {Fabre}, {Fabrizio},
  {Faigler}, {Fedorets}, {Fernique}, {Fienga}, {Figueras}, {Fournier},
  {Fouron}, {Fragkoudi}, {Gai}, {Garcia-Gutierrez}, {Garcia-Reinaldos},
  {Garc{\'\i}a-Torres}, {Garofalo}, {Gavel}, {Gavras}, {Gerlach}, {Geyer},
  {Giacobbe}, {Gilmore}, {Girona}, {Giuffrida}, {Gomel}, {Gomez},
  {Gonz{\'a}lez-N{\'u}{\~n}ez}, {Gonz{\'a}lez-Santamar{\'\i}a},
  {Gonz{\'a}lez-Vidal}, {Granvik}, {Guillout}, {Guiraud},
  {Guti{\'e}rrez-S{\'a}nchez}, {Guy}, {Hatzidimitriou}, {Hauser}, {Haywood},
  {Helmer}, {Helmi}, {Sarmiento}, {Hidalgo}, {Hilger}, {H{\l}adczuk}, {Hobbs},
  {Holland}, {Huckle}, {Jardine}, {Jasniewicz}, {Jean-Antoine Piccolo},
  {Jim{\'e}nez-Arranz}, {Jorissen}, {Juaristi Campillo}, {Julbe}, {Karbevska},
  {Kervella}, {Khanna}, {Kontizas}, {Kordopatis}, {Korn}, {K{\'o}sp{\'a}l},
  {Kostrzewa-Rutkowska}, {Kruszy{\'n}ska}, {Kun}, {Laizeau}, {Lambert},
  {Lanza}, {Lasne}, {Le Campion}, {Lebreton}, {Lebzelter}, {Leccia}, {Leclerc},
  {Lecoeur-Taibi}, {Liao}, {Licata}, {Lindstr{\o}m}, {Lister}, {Livanou},
  {Lobel}, {Lorca}, {Loup}, {Madrero Pardo}, {Magdaleno Romeo}, {Managau},
  {Mann}, {Manteiga}, {Marchant}, {Marconi}, {Marcos}, {Marcos Santos},
  {Mar{\'\i}n Pina}, {Marinoni}, {Marocco}, {Marshall}, {Martin Polo},
  {Mart{\'\i}n-Fleitas}, {Marton}, {Mary}, {Masip}, {Massari},
  {Mastrobuono-Battisti}, {Mazeh}, {McMillan}, {Messina}, {Michalik}, {Millar},
  {Mints}, {Molina}, {Molinaro}, {Moln{\'a}r}, {Monari}, {Mongui{\'o}},
  {Montegriffo}, {Montero}, {Mor}, {Mora}, {Morbidelli}, {Morel}, {Morris},
  {Muraveva}, {Murphy}, {Musella}, {Nagy}, {Noval}, {Oca{\~n}a}, {Ogden},
  {Ordenovic}, {Osinde}, {Pagani}, {Pagano}, {Palaversa}, {Palicio},
  {Pallas-Quintela}, {Panahi}, {Payne-Wardenaar}, {Pe{\~n}alosa Esteller},
  {Penttil{\"a}}, {Pichon}, {Piersimoni}, {Pineau}, {Plachy}, {Plum}, {Poggio},
  {Pr{\v{s}}a}, {Pulone}, {Racero}, {Ragaini}, {Rainer}, {Raiteri}, {Rambaux},
  {Ramos}, {Ramos-Lerate}, {Re Fiorentin}, {Regibo}, {Richards}, {Rios Diaz},
  {Ripepi}, {Riva}, {Rix}, {Rixon}, {Robichon}, {Robin}, {Robin}, {Roelens},
  {Rogues}, {Rohrbasser}, {Romero-G{\'o}mez}, {Rowell}, {Royer}, {Ruz Mieres},
  {Rybicki}, {Sadowski}, {S{\'a}ez N{\'u}{\~n}ez}, {Sagrist{\`a} Sell{\'e}s},
  {Sahlmann}, {Salguero}, {Samaras}, {Sanchez Gimenez}, {Sanna},
  {Santove{\~n}a}, {Sarasso}, {Schultheis}, {Sciacca}, {Segol}, {Segovia},
  {S{\'e}gransan}, {Semeux}, {Shahaf}, {Siddiqui}, {Siebert}, {Siltala},
  {Silvelo}, {Slezak}, {Slezak}, {Smart}, {Snaith}, {Solano}, {Solitro},
  {Souami}, {Souchay}, {Spagna}, {Spina}, {Spoto}, {Steele},
  {Steidelm{\"u}ller}, {Stephenson}, {S{\"u}veges}, {Surdej}, {Szabados},
  {Szegedi-Elek}, {Taris}, {Taylor}, {Teixeira}, {Tolomei}, {Tonello}, {Torra},
  {Torra}, {Torralba Elipe}, {Trabucchi}, {Tsounis}, {Turon}, {Ulla}, {Unger},
  {Vaillant}, {van Dillen}, {van Reeven}, {Vanel}, {Vecchiato}, {Viala},
  {Vicente}, {Voutsinas}, {Weiler}, {Wevers}, {Wyrzykowski}, {Yoldas}, {Yvard},
  {Zhao}, {Zorec}, {Zucker}, \& {Zwitter}}]{gaia23}
{Gaia Collaboration}, {Vallenari}, A., {Brown}, A.~G.~A., {et~al.} 2023, \aap,
  674, A1

\bibitem[{{Gilbertson} {et~al.}(2020){Gilbertson}, {Ford}, \&
  {Dumusque}}]{gilbertson20}
{Gilbertson}, C., {Ford}, E.~B., \& {Dumusque}, X. 2020, Research Notes of the
  American Astronomical Society, 4, 59

\bibitem[{{Giles} {et~al.}(2017){Giles}, {Collier Cameron}, \&
  {Haywood}}]{giles17}
{Giles}, H. A.~C., {Collier Cameron}, A., \& {Haywood}, R.~D. 2017, \mnras,
  472, 1618

\bibitem[{{Goldreich} \& {Soter}(1966)}]{goldreich66}
{Goldreich}, P., \& {Soter}, S. 1966, \icarus, 5, 375

\bibitem[{{Hadden} \& {Lithwick}(2017)}]{hadden17}
{Hadden}, S., \& {Lithwick}, Y. 2017, \aj, 154, 5

\bibitem[{{Halverson} {et~al.}(2016){Halverson}, {Terrien}, {Mahadevan}, {Roy},
  {Bender}, {Stef{\'a}nsson}, {Monson}, {Levi}, {Hearty}, {Blake}, {McElwain},
  {Schwab}, {Ramsey}, {Wright}, {Wang}, {Gong}, \& {Roberston}}]{halverson16}
{Halverson}, S., {Terrien}, R., {Mahadevan}, S., {et~al.} 2016, in Society of
  Photo-Optical Instrumentation Engineers (SPIE) Conference Series, Vol. 9908,
  Ground-based and Airborne Instrumentation for Astronomy VI, ed. C.~J.
  {Evans}, L.~{Simard}, \& H.~{Takami}, 99086P

\bibitem[{{Harada} {et~al.}(2024){Harada}, {Dressing}, {Kane}, \& {Adami
  Ardestani}}]{harada24}
{Harada}, C.~K., {Dressing}, C.~D., {Kane}, S.~R., \& {Adami Ardestani}, B.
  2024, arXiv e-prints, arXiv:2401.03047

\bibitem[{{Haywood} {et~al.}(2014){Haywood}, {Collier Cameron}, {Queloz},
  {Barros}, {Deleuil}, {Fares}, {Gillon}, {Lanza}, {Lovis}, {Moutou}, {Pepe},
  {Pollacco}, {Santerne}, {S{\'e}gransan}, \& {Unruh}}]{haywood14}
{Haywood}, R.~D., {Collier Cameron}, A., {Queloz}, D., {et~al.} 2014, \mnras,
  443, 2517

\bibitem[{{Hippke} \& {Heller}(2019)}]{hippke19}
{Hippke}, M., \& {Heller}, R. 2019, \aap, 623, A39

\bibitem[{{Hoffman} \& {Gelman}(2011)}]{hoffman11}
{Hoffman}, M.~D., \& {Gelman}, A. 2011, arXiv e-prints, arXiv:1111.4246

\bibitem[{{Holcomb} {et~al.}(2022){Holcomb}, {Robertson}, {Hartigan},
  {Oelkers}, \& {Robinson}}]{holcomb22}
{Holcomb}, R.~J., {Robertson}, P., {Hartigan}, P., {Oelkers}, R.~J., \&
  {Robinson}, C. 2022, \apj, 936, 138

\bibitem[{{Howard} {et~al.}(2010){Howard}, {Johnson}, {Marcy}, {Fischer},
  {Wright}, {Bernat}, {Henry}, {Peek}, {Isaacson}, {Apps}, {Endl}, {Cochran},
  {Valenti}, {Anderson}, \& {Piskunov}}]{howard10}
{Howard}, A.~W., {Johnson}, J.~A., {Marcy}, G.~W., {et~al.} 2010, \apj, 721,
  1467

\bibitem[{{Howard} {et~al.}(2012){Howard}, {Marcy}, {Bryson}, {Jenkins},
  {Rowe}, {Batalha}, {Borucki}, {Koch}, {Dunham}, {Gautier}, {Van Cleve},
  {Cochran}, {Latham}, {Lissauer}, {Torres}, {Brown}, {Gilliland}, {Buchhave},
  {Caldwell}, {Christensen-Dalsgaard}, {Ciardi}, {Fressin}, {Haas}, {Howell},
  {Kjeldsen}, {Seager}, {Rogers}, {Sasselov}, {Steffen}, {Basri},
  {Charbonneau}, {Christiansen}, {Clarke}, {Dupree}, {Fabrycky}, {Fischer},
  {Ford}, {Fortney}, {Tarter}, {Girouard}, {Holman}, {Johnson}, {Klaus},
  {Machalek}, {Moorhead}, {Morehead}, {Ragozzine}, {Tenenbaum}, {Twicken},
  {Quinn}, {Isaacson}, {Shporer}, {Lucas}, {Walkowicz}, {Welsh}, {Boss},
  {Devore}, {Gould}, {Smith}, {Morris}, {Prsa}, {Morton}, {Still}, {Thompson},
  {Mullally}, {Endl}, \& {MacQueen}}]{howard12}
{Howard}, A.~W., {Marcy}, G.~W., {Bryson}, S.~T., {et~al.} 2012, \apjs, 201, 15

\bibitem[{{Howell} {et~al.}(2012){Howell}, {Rowe}, {Bryson}, {Quinn}, {Marcy},
  {Isaacson}, {Ciardi}, {Chaplin}, {Metcalfe}, {Monteiro}, {Appourchaux},
  {Basu}, {Creevey}, {Gilliland}, {Quirion}, {Stello}, {Kjeldsen},
  {Christensen-Dalsgaard}, {Elsworth}, {Garc{\'\i}a}, {Houdek}, {Karoff},
  {Molenda-{\.Z}akowicz}, {Thompson}, {Verner}, {Torres}, {Fressin}, {Crepp},
  {Adams}, {Dupree}, {Sasselov}, {Dressing}, {Borucki}, {Koch}, {Lissauer},
  {Latham}, {Buchhave}, {Gautier}, {Everett}, {Horch}, {Batalha}, {Dunham},
  {Szkody}, {Silva}, {Mighell}, {Holberg}, {Ballot}, {Bedding}, {Bruntt},
  {Campante}, {Handberg}, {Hekker}, {Huber}, {Mathur}, {Mosser}, {R{\'e}gulo},
  {White}, {Christiansen}, {Middour}, {Haas}, {Hall}, {Jenkins}, {McCaulif},
  {Fanelli}, {Kulesa}, {McCarthy}, \& {Henze}}]{howell12}
{Howell}, S.~B., {Rowe}, J.~F., {Bryson}, S.~T., {et~al.} 2012, \apj, 746, 123

\bibitem[{{Jenkins} {et~al.}(2010){Jenkins}, {Caldwell}, {Chandrasekaran},
  {Twicken}, {Bryson}, {Quintana}, {Clarke}, {Li}, {Allen}, {Tenenbaum}, {Wu},
  {Klaus}, {Middour}, {Cote}, {McCauliff}, {Girouard}, {Gunter}, {Wohler},
  {Sommers}, {Hall}, {Uddin}, {Wu}, {Bhavsar}, {Van Cleve}, {Pletcher},
  {Dotson}, {Haas}, {Gilliland}, {Koch}, \& {Borucki}}]{jenkins10}
{Jenkins}, J.~M., {Caldwell}, D.~A., {Chandrasekaran}, H., {et~al.} 2010,
  \apjl, 713, L87

\bibitem[{{Jenkins} {et~al.}(2016){Jenkins}, {Twicken}, {McCauliff},
  {Campbell}, {Sanderfer}, {Lung}, {Mansouri-Samani}, {Girouard}, {Tenenbaum},
  {Klaus}, {Smith}, {Caldwell}, {Chacon}, {Henze}, {Heiges}, {Latham},
  {Morgan}, {Swade}, {Rinehart}, \& {Vanderspek}}]{jenkins16}
{Jenkins}, J.~M., {Twicken}, J.~D., {McCauliff}, S., {et~al.} 2016, in Society
  of Photo-Optical Instrumentation Engineers (SPIE) Conference Series, Vol.
  9913, Software and Cyberinfrastructure for Astronomy IV, ed. G.~{Chiozzi} \&
  J.~C. {Guzman}, 99133E

\bibitem[{Kass \& Raftery(1995)}]{kass95}
Kass, R.~E., \& Raftery, A.~E. 1995, Journal of the american statistical
  association, 90, 773

\bibitem[{{Kempton} {et~al.}(2018){Kempton}, {Bean}, {Louie}, {Deming}, {Koll},
  {Mansfield}, {Christiansen}, {L{\'o}pez-Morales}, {Swain}, {Zellem},
  {Ballard}, {Barclay}, {Barstow}, {Batalha}, {Beatty}, {Berta-Thompson},
  {Birkby}, {Buchhave}, {Charbonneau}, {Cowan}, {Crossfield}, {de Val-Borro},
  {Doyon}, {Dragomir}, {Gaidos}, {Heng}, {Hu}, {Kane}, {Kreidberg}, {Mallonn},
  {Morley}, {Narita}, {Nascimbeni}, {Pall{\'e}}, {Quintana}, {Rauscher},
  {Seager}, {Shkolnik}, {Sing}, {Sozzetti}, {Stassun}, {Valenti}, \& {von
  Essen}}]{kempton18}
{Kempton}, E. M.~R., {Bean}, J.~L., {Louie}, D.~R., {et~al.} 2018, \pasp, 130,
  114401

\bibitem[{{Kossakowski} {et~al.}(2021){Kossakowski}, {Kemmer}, {Bluhm},
  {Stock}, {Caballero}, {B{\'e}jar}, {Guill{\'e}n}, {Lodieu}, {Collins},
  {Oshagh}, {Schlecker}, {Espinoza}, {Pall{\'e}}, {Henning}, {Kreidberg},
  {K{\"u}rster}, {Amado}, {Anderson}, {Morales}, {Cartwright}, {Charbonneau},
  {Chaturvedi}, {Cifuentes}, {Conti}, {Cort{\'e}s-Contreras}, {Dreizler},
  {Galad{\'\i}-Enr{\'\i}quez}, {Guerra}, {Hart}, {Hellier}, {Henze}, {Herrero},
  {Jeffers}, {Jenkins}, {Jensen}, {Kaminski}, {Kielkopf}, {Kunimoto},
  {Lafarga}, {Latham}, {Lillo-Box}, {Luque}, {Molaverdikhani}, {Montes},
  {Morello}, {Morgan}, {Nowak}, {Pavlov}, {Perger}, {Quintana}, {Quirrenbach},
  {Reffert}, {Reiners}, {Ricker}, {Ribas}, {Rodr{\'\i}guez L{\'o}pez},
  {Zapatero Osorio}, {Seager}, {Sch{\"o}fer}, {Schweitzer}, {Trifonov},
  {Vanaverbeke}, {Vanderspek}, {West}, {Winn}, \&
  {Zechmeister}}]{kossakowski21}
{Kossakowski}, D., {Kemmer}, J., {Bluhm}, P., {et~al.} 2021, \aap, 656, A124

\bibitem[{Kumar {et~al.}(2019)Kumar, Carroll, Hartikainen, \&
  Martin}]{exoplanet:arviz}
Kumar, R., Carroll, C., Hartikainen, A., \& Martin, O.~A. 2019, The Journal of
  Open Source Software, doi:10.21105/joss.01143.
\newblock \url{http://joss.theoj.org/papers/10.21105/joss.01143}

\bibitem[{{Li} {et~al.}(2021){Li}, {Hildebrandt}, {Kane}, {Zimmerman},
  {Girard}, {Gonzalez-Quiles}, \& {Turnbull}}]{li21}
{Li}, Z., {Hildebrandt}, S.~R., {Kane}, S.~R., {et~al.} 2021, \aj, 162, 9

\bibitem[{{Liddle}(2007)}]{liddle07}
{Liddle}, A.~R. 2007, \mnras, 377, L74

\bibitem[{{Lightkurve Collaboration} {et~al.}(2018){Lightkurve Collaboration},
  {Cardoso}, {Hedges}, {Gully-Santiago}, {Saunders}, {Cody}, {Barclay}, {Hall},
  {Sagear}, {Turtelboom}, {Zhang}, {Tzanidakis}, {Mighell}, {Coughlin}, {Bell},
  {Berta-Thompson}, {Williams}, {Dotson}, \& {Barentsen}}]{lightkurve18}
{Lightkurve Collaboration}, {Cardoso}, J.~V.~d.~M., {Hedges}, C., {et~al.}
  2018, {Lightkurve: Kepler and TESS time series analysis in Python},
  Astrophysics Source Code Library, , , ascl:1812.013

\bibitem[{{Lindegren} {et~al.}(2018){Lindegren}, {Hern{\'a}ndez}, {Bombrun},
  {Klioner}, {Bastian}, {Ramos-Lerate}, {de Torres}, {Steidelm{\"u}ller},
  {Stephenson}, {Hobbs}, {Lammers}, {Biermann}, {Geyer}, {Hilger}, {Michalik},
  {Stampa}, {McMillan}, {Casta{\~n}eda}, {Clotet}, {Comoretto}, {Davidson},
  {Fabricius}, {Gracia}, {Hambly}, {Hutton}, {Mora}, {Portell}, {van Leeuwen},
  {Abbas}, {Abreu}, {Altmann}, {Andrei}, {Anglada}, {Balaguer-N{\'u}{\~n}ez},
  {Barache}, {Becciani}, {Bertone}, {Bianchi}, {Bouquillon}, {Bourda},
  {Br{\"u}semeister}, {Bucciarelli}, {Busonero}, {Buzzi}, {Cancelliere},
  {Carlucci}, {Charlot}, {Cheek}, {Crosta}, {Crowley}, {de Bruijne}, {de
  Felice}, {Drimmel}, {Esquej}, {Fienga}, {Fraile}, {Gai}, {Garralda},
  {Gonz{\'a}lez-Vidal}, {Guerra}, {Hauser}, {Hofmann}, {Holl}, {Jordan},
  {Lattanzi}, {Lenhardt}, {Liao}, {Licata}, {Lister}, {L{\"o}ffler},
  {Marchant}, {Martin-Fleitas}, {Messineo}, {Mignard}, {Morbidelli}, {Poggio},
  {Riva}, {Rowell}, {Salguero}, {Sarasso}, {Sciacca}, {Siddiqui}, {Smart},
  {Spagna}, {Steele}, {Taris}, {Torra}, {van Elteren}, {van Reeven}, \&
  {Vecchiato}}]{lindegren18}
{Lindegren}, L., {Hern{\'a}ndez}, J., {Bombrun}, A., {et~al.} 2018, \aap, 616,
  A2

\bibitem[{{L{\'o}pez-Morales} {et~al.}(2016){L{\'o}pez-Morales}, {Haywood},
  {Coughlin}, {Zeng}, {Buchhave}, {Giles}, {Affer}, {Bonomo}, {Charbonneau},
  {Collier Cameron}, {Consentino}, {Dressing}, {Dumusque}, {Figueira},
  {Fiorenzano}, {Harutyunyan}, {Johnson}, {Latham}, {Lopez}, {Lovis},
  {Malavolta}, {Mayor}, {Micela}, {Molinari}, {Mortier}, {Motalebi},
  {Nascimbeni}, {Pepe}, {Phillips}, {Piotto}, {Pollacco}, {Queloz}, {Rice},
  {Sasselov}, {Segransan}, {Sozzetti}, {Udry}, {Vanderburg}, \&
  {Watson}}]{lopezmorales16}
{L{\'o}pez-Morales}, M., {Haywood}, R.~D., {Coughlin}, J.~L., {et~al.} 2016,
  \aj, 152, 204

\bibitem[{{Lubin} {et~al.}(2021){Lubin}, {Robertson}, {Stefansson}, {Ninan},
  {Mahadevan}, {Endl}, {Ford}, {Wright}, {Beard}, {Bender}, {Cochran},
  {Diddams}, {Fredrick}, {Halverson}, {Kanodia}, {Metcalf}, {Ramsey}, {Roy},
  {Schwab}, \& {Terrien}}]{lubin21}
{Lubin}, J., {Robertson}, P., {Stefansson}, G., {et~al.} 2021, \aj, 162, 61

\bibitem[{{Lubin} {et~al.}(2022){Lubin}, {Van Zandt}, {Holcomb}, {Weiss},
  {Petigura}, {Robertson}, {Akana Murphy}, {Scarsdale}, {Batygin}, {Polanski},
  {Batalha}, {Crossfield}, {Dressing}, {Fulton}, {Howard}, {Huber}, {Isaacson},
  {Kane}, {Roy}, {Beard}, {Blunt}, {Chontos}, {Dai}, {Dalba}, {Gary},
  {Giacalone}, {Hill}, {Mayo}, {Mo{\v{c}}nik}, {Kosiarek}, {Rice}, {Rubenzahl},
  {Latham}, {Seager}, {Winn}, \& {Gary}}]{lubin22}
{Lubin}, J., {Van Zandt}, J., {Holcomb}, R., {et~al.} 2022, \aj, 163, 101

\bibitem[{{Luger} {et~al.}(2019){Luger}, {Agol}, {Foreman-Mackey}, {Fleming},
  {Lustig-Yaeger}, \& {Deitrick}}]{exoplanet:luger18}
{Luger}, R., {Agol}, E., {Foreman-Mackey}, D., {et~al.} 2019, \aj, 157, 64

\bibitem[{{Luque} \& {Pall{\'e}}(2022)}]{luque22}
{Luque}, R., \& {Pall{\'e}}, E. 2022, Science, 377, 1211

\bibitem[{{Luque} {et~al.}(2019){Luque}, {Pall{\'e}}, {Kossakowski},
  {Dreizler}, {Kemmer}, {Espinoza}, {Burt}, {Anglada-Escud{\'e}}, {B{\'e}jar},
  {Caballero}, {Collins}, {Collins}, {Cort{\'e}s-Contreras},
  {D{\'\i}ez-Alonso}, {Feng}, {Hatzes}, {Hellier}, {Henning}, {Jeffers},
  {Kaltenegger}, {K{\"u}rster}, {Madden}, {Molaverdikhani}, {Montes}, {Narita},
  {Nowak}, {Ofir}, {Oshagh}, {Parviainen}, {Quirrenbach}, {Reffert}, {Reiners},
  {Rodr{\'\i}guez-L{\'o}pez}, {Schlecker}, {Stock}, {Trifonov}, {Winn},
  {Zapatero Osorio}, {Zechmeister}, {Amado}, {Anderson}, {Batalha}, {Bauer},
  {Bluhm}, {Burke}, {Butler}, {Caldwell}, {Chen}, {Crane}, {Dragomir},
  {Dressing}, {Dynes}, {Jenkins}, {Kaminski}, {Klahr}, {Kotani}, {Lafarga},
  {Latham}, {Lewin}, {McDermott}, {Monta{\~n}{\'e}s-Rodr{\'\i}guez}, {Morales},
  {Murgas}, {Nagel}, {Pedraz}, {Ribas}, {Ricker}, {Rowden}, {Seager},
  {Shectman}, {Tamura}, {Teske}, {Twicken}, {Vanderspeck}, {Wang}, \&
  {Wohler}}]{luque19}
{Luque}, R., {Pall{\'e}}, E., {Kossakowski}, D., {et~al.} 2019, \aap, 628, A39

\bibitem[{{Luri} {et~al.}(2018){Luri}, {Brown}, {Sarro}, {Arenou},
  {Bailer-Jones}, {Castro-Ginard}, {de Bruijne}, {Prusti}, {Babusiaux}, \&
  {Delgado}}]{luri18}
{Luri}, X., {Brown}, A.~G.~A., {Sarro}, L.~M., {et~al.} 2018, \aap, 616, A9

\bibitem[{{Mayor} \& {Queloz}(1995)}]{mayor95}
{Mayor}, M., \& {Queloz}, D. 1995, \nat, 378, 355

\bibitem[{{McArthur} {et~al.}(2004){McArthur}, {Endl}, {Cochran}, {Benedict},
  {Fischer}, {Marcy}, {Butler}, {Naef}, {Mayor}, {Queloz}, {Udry}, \&
  {Harrison}}]{mcarthur04}
{McArthur}, B.~E., {Endl}, M., {Cochran}, W.~D., {et~al.} 2004, \apjl, 614, L81

\bibitem[{{McQuillan} {et~al.}(2014){McQuillan}, {Mazeh}, \&
  {Aigrain}}]{mcquillan14}
{McQuillan}, A., {Mazeh}, T., \& {Aigrain}, S. 2014, \apjs, 211, 24

\bibitem[{{Murray} \& {Dermott}(1999)}]{murray99}
{Murray}, C.~D., \& {Dermott}, S.~F. 1999, {Solar system dynamics}

\bibitem[{{National Academies of Sciences} \& Medicine(2021)}]{decadal20}
{National Academies of Sciences}, E., \& Medicine. 2021, {Pathways to Discovery
  in Astronomy and Astrophysics for the 2020s}, doi:10.17226/26141

\bibitem[{{Nelson} {et~al.}(2020){Nelson}, {Ford}, {Buchner}, {Cloutier},
  {D{\'\i}az}, {Faria}, {Hara}, {Rajpaul}, \& {Rukdee}}]{nelson20}
{Nelson}, B.~E., {Ford}, E.~B., {Buchner}, J., {et~al.} 2020, \aj, 159, 73

\bibitem[{{Rauer} {et~al.}(2014){Rauer}, {Catala}, {Aerts}, {Appourchaux},
  {Benz}, {Brandeker}, {Christensen-Dalsgaard}, {Deleuil}, {Gizon}, {Goupil},
  {G{\"u}del}, {Janot-Pacheco}, {Mas-Hesse}, {Pagano}, {Piotto}, {Pollacco},
  {Santos}, {Smith}, {Su{\'a}rez}, {Szab{\'o}}, {Udry}, {Adibekyan}, {Alibert},
  {Almenara}, {Amaro-Seoane}, {Eiff}, {Asplund}, {Antonello}, {Barnes},
  {Baudin}, {Belkacem}, {Bergemann}, {Bihain}, {Birch}, {Bonfils}, {Boisse},
  {Bonomo}, {Borsa}, {Brand{\~a}o}, {Brocato}, {Brun}, {Burleigh}, {Burston},
  {Cabrera}, {Cassisi}, {Chaplin}, {Charpinet}, {Chiappini}, {Church},
  {Csizmadia}, {Cunha}, {Damasso}, {Davies}, {Deeg}, {D{\'\i}az}, {Dreizler},
  {Dreyer}, {Eggenberger}, {Ehrenreich}, {Eigm{\"u}ller}, {Erikson}, {Farmer},
  {Feltzing}, {de Oliveira Fialho}, {Figueira}, {Forveille}, {Fridlund},
  {Garc{\'\i}a}, {Giommi}, {Giuffrida}, {Godolt}, {Gomes da Silva}, {Granzer},
  {Grenfell}, {Grotsch-Noels}, {G{\"u}nther}, {Haswell}, {Hatzes},
  {H{\'e}brard}, {Hekker}, {Helled}, {Heng}, {Jenkins}, {Johansen},
  {Khodachenko}, {Kislyakova}, {Kley}, {Kolb}, {Krivova}, {Kupka}, {Lammer},
  {Lanza}, {Lebreton}, {Magrin}, {Marcos-Arenal}, {Marrese}, {Marques},
  {Martins}, {Mathis}, {Mathur}, {Messina}, {Miglio}, {Montalban}, {Montalto},
  {Monteiro}, {Moradi}, {Moravveji}, {Mordasini}, {Morel}, {Mortier},
  {Nascimbeni}, {Nelson}, {Nielsen}, {Noack}, {Norton}, {Ofir}, {Oshagh},
  {Ouazzani}, {P{\'a}pics}, {Parro}, {Petit}, {Plez}, {Poretti}, {Quirrenbach},
  {Ragazzoni}, {Raimondo}, {Rainer}, {Reese}, {Redmer}, {Reffert},
  {Rojas-Ayala}, {Roxburgh}, {Salmon}, {Santerne}, {Schneider}, {Schou},
  {Schuh}, {Schunker}, {Silva-Valio}, {Silvotti}, {Skillen}, {Snellen}, {Sohl},
  {Sousa}, {Sozzetti}, {Stello}, {Strassmeier}, {{\v{S}}vanda}, {Szab{\'o}},
  {Tkachenko}, {Valencia}, {Van Grootel}, {Vauclair}, {Ventura}, {Wagner},
  {Walton}, {Weingrill}, {Werner}, {Wheatley}, \& {Zwintz}}]{rauer14}
{Rauer}, H., {Catala}, C., {Aerts}, C., {et~al.} 2014, Experimental Astronomy,
  38, 249

\bibitem[{{Ricker} {et~al.}(2015){Ricker}, {Winn}, {Vanderspek}, {Latham},
  {Bakos}, {Bean}, {Berta-Thompson}, {Brown}, {Buchhave}, {Butler}, {Butler},
  {Chaplin}, {Charbonneau}, {Christensen-Dalsgaard}, {Clampin}, {Deming},
  {Doty}, {De Lee}, {Dressing}, {Dunham}, {Endl}, {Fressin}, {Ge}, {Henning},
  {Holman}, {Howard}, {Ida}, {Jenkins}, {Jernigan}, {Johnson}, {Kaltenegger},
  {Kawai}, {Kjeldsen}, {Laughlin}, {Levine}, {Lin}, {Lissauer}, {MacQueen},
  {Marcy}, {McCullough}, {Morton}, {Narita}, {Paegert}, {Palle}, {Pepe},
  {Pepper}, {Quirrenbach}, {Rinehart}, {Sasselov}, {Sato}, {Seager},
  {Sozzetti}, {Stassun}, {Sullivan}, {Szentgyorgyi}, {Torres}, {Udry}, \&
  {Villasenor}}]{ricker15}
{Ricker}, G.~R., {Winn}, J.~N., {Vanderspek}, R., {et~al.} 2015, Journal of
  Astronomical Telescopes, Instruments, and Systems, 1, 014003

\bibitem[{Salvatier {et~al.}(2016)Salvatier, Wiecki, \&
  Fonnesbeck}]{exoplanet:pymc3}
Salvatier, J., Wiecki, T.~V., \& Fonnesbeck, C. 2016, PeerJ Computer Science,
  2, e55

\bibitem[{{Schwab} {et~al.}(2016){Schwab}, {Rakich}, {Gong}, {Mahadevan},
  {Halverson}, {Roy}, {Terrien}, {Robertson}, {Hearty}, {Levi}, {Monson},
  {Wright}, {McElwain}, {Bender}, {Blake}, {St{\"u}rmer}, {Gurevich},
  {Chakraborty}, \& {Ramsey}}]{schwab16}
{Schwab}, C., {Rakich}, A., {Gong}, Q., {et~al.} 2016, in Society of
  Photo-Optical Instrumentation Engineers (SPIE) Conference Series, Vol. 9908,
  Ground-based and Airborne Instrumentation for Astronomy VI, ed. C.~J.
  {Evans}, L.~{Simard}, \& H.~{Takami}, 99087H

\bibitem[{{Silva Aguirre} {et~al.}(2015){Silva Aguirre}, {Davies}, {Basu},
  {Christensen-Dalsgaard}, {Creevey}, {Metcalfe}, {Bedding}, {Casagrande},
  {Handberg}, {Lund}, {Nissen}, {Chaplin}, {Huber}, {Serenelli}, {Stello}, {Van
  Eylen}, {Campante}, {Elsworth}, {Gilliland}, {Hekker}, {Karoff}, {Kawaler},
  {Kjeldsen}, \& {Lundkvist}}]{aguirre15}
{Silva Aguirre}, V., {Davies}, G.~R., {Basu}, S., {et~al.} 2015, \mnras, 452,
  2127

\bibitem[{{Sozzetti}(2023)}]{sozzetti23}
{Sozzetti}, A. 2023, \aap, 670, L17

\bibitem[{{Stapelfeldt}(2006)}]{stapelfeldt06}
{Stapelfeldt}, K.~R. 2006, in The Scientific Requirements for Extremely Large
  Telescopes, ed. P.~{Whitelock}, M.~{Dennefeld}, \& B.~{Leibundgut}, Vol. 232,
  149--158

\bibitem[{{Stassun} {et~al.}(2018){Stassun}, {Oelkers}, {Pepper}, {Paegert},
  {De Lee}, {Torres}, {Latham}, {Charpinet}, {Dressing}, {Huber}, {Kane},
  {L{\'e}pine}, {Mann}, {Muirhead}, {Rojas-Ayala}, {Silvotti}, {Fleming},
  {Levine}, \& {Plavchan}}]{stassun18}
{Stassun}, K.~G., {Oelkers}, R.~J., {Pepper}, J., {et~al.} 2018, \aj, 156, 102

\bibitem[{{Stef{\`a}nsson} {et~al.}(2022){Stef{\`a}nsson}, {Mahadevan},
  {Petrovich}, {Winn}, {Kanodia}, {Millholland}, {Maney}, {Ca{\~n}as},
  {Wisniewski}, {Robertson}, {Ninan}, {Ford}, {Bender}, {Blake}, {Cegla},
  {Cochran}, {Diddams}, {Dong}, {Endl}, {Fredrick}, {Halverson}, {Hearty},
  {Hebb}, {Hirano}, {Lin}, {Logsdon}, {Lubar}, {McElwain}, {Metcalf}, {Monson},
  {Rajagopal}, {Ramsey}, {Roy}, {Schwab}, {Schweiker}, {Terrien}, \&
  {Wright}}]{stefansson22}
{Stef{\`a}nsson}, G., {Mahadevan}, S., {Petrovich}, C., {et~al.} 2022, \apjl,
  931, L15

\bibitem[{{Theano Development Team}(2016)}]{exoplanet:theano}
{Theano Development Team}. 2016, arXiv e-prints, abs/1605.02688.
\newblock \url{http://arxiv.org/abs/1605.02688}

\bibitem[{{Tinney} {et~al.}(2005){Tinney}, {Butler}, {Marcy}, {Jones}, {Penny},
  {McCarthy}, {Carter}, \& {Fischer}}]{tinney05}
{Tinney}, C.~G., {Butler}, R.~P., {Marcy}, G.~W., {et~al.} 2005, \apj, 623,
  1171

\bibitem[{{Trotta}(2008)}]{trotta08}
{Trotta}, R. 2008, Contemporary Physics, 49, 71

\bibitem[{{Vogt} {et~al.}(1994){Vogt}, {Allen}, {Bigelow}, {Bresee}, {Brown},
  {Cantrall}, {Conrad}, {Couture}, {Delaney}, {Epps}, {Hilyard}, {Hilyard},
  {Horn}, {Jern}, {Kanto}, {Keane}, {Kibrick}, {Lewis}, {Osborne},
  {Pardeilhan}, {Pfister}, {Ricketts}, {Robinson}, {Stover}, {Tucker}, {Ward},
  \& {Wei}}]{vogt94}
{Vogt}, S.~S., {Allen}, S.~L., {Bigelow}, B.~C., {et~al.} 1994, in Society of
  Photo-Optical Instrumentation Engineers (SPIE) Conference Series, Vol. 2198,
  Instrumentation in Astronomy VIII, ed. D.~L. {Crawford} \& E.~R. {Craine},
  362

\bibitem[{{Wang} {et~al.}(2020){Wang}, {Ginzburg}, {Ren}, {Wallack}, {Gao},
  {Mawet}, {Bond}, {Cetre}, {Wizinowich}, {De Rosa}, {Ruane}, {Liu}, {Absil},
  {Alvarez}, {Baranec}, {Choquet}, {Chun}, {Defr{\`e}re}, {Delorme},
  {Duch{\^e}ne}, {Forsberg}, {Ghez}, {Guyon}, {Hall}, {Huby}, {Jolivet},
  {Jensen-Clem}, {Jovanovic}, {Karlsson}, {Lilley}, {Matthews}, {M{\'e}nard},
  {Meshkat}, {Millar-Blanchaer}, {Ngo}, {Orban de Xivry}, {Pinte}, {Ragland},
  {Serabyn}, {Catal{\'a}n}, {Wang}, {Wetherell}, {Williams}, {Ygouf}, \&
  {Zuckerman}}]{wang20}
{Wang}, J.~J., {Ginzburg}, S., {Ren}, B., {et~al.} 2020, \aj, 159, 263

\bibitem[{{Wright} \& {Robertson}(2017)}]{wright17}
{Wright}, J.~T., \& {Robertson}, P. 2017, Research Notes of the American
  Astronomical Society, 1, 51

\bibitem[{{Wu} {et~al.}(2010){Wu}, {Twicken}, {Tenenbaum}, {Clarke}, {Li},
  {Quintana}, {Allen}, {Chandrasekaran}, {Jenkins}, {Caldwell}, {Wohler},
  {Girouard}, {McCauliff}, {Cote}, \& {Klaus}}]{wu10}
{Wu}, H., {Twicken}, J.~D., {Tenenbaum}, P., {et~al.} 2010, in Society of
  Photo-Optical Instrumentation Engineers (SPIE) Conference Series, Vol. 7740,
  Software and Cyberinfrastructure for Astronomy, ed. N.~M. {Radziwill} \&
  A.~{Bridger}, 774019

\bibitem[{{Zechmeister} {et~al.}(2018){Zechmeister}, {Reiners}, {Amado},
  {Azzaro}, {Bauer}, {B{\'e}jar}, {Caballero}, {Guenther}, {Hagen}, {Jeffers},
  {Kaminski}, {K{\"u}rster}, {Launhardt}, {Montes}, {Morales}, {Quirrenbach},
  {Reffert}, {Ribas}, {Seifert}, {Tal-Or}, \& {Wolthoff}}]{zechmeister18}
{Zechmeister}, M., {Reiners}, A., {Amado}, P.~J., {et~al.} 2018, \aap, 609, A12

\bibitem[{{Zeng} {et~al.}(2019){Zeng}, {Jacobsen}, {Sasselov}, {Petaev},
  {Vanderburg}, {Lopez-Morales}, {Perez-Mercader}, {Mattsson}, {Li}, {Heising},
  {Bonomo}, {Damasso}, {Berger}, {Cao}, {Levi}, \& {Wordsworth}}]{zeng19}
{Zeng}, L., {Jacobsen}, S.~B., {Sasselov}, D.~D., {et~al.} 2019, Proceedings of
  the National Academy of Science, 116, 9723

\end{thebibliography}

\appendix 

We include the NEID and HIRES RV data for Kepler-21 in Tables \ref{tab:neid_data} and \ref{tab:hires_data}. The HARPS-N data is available in \cite{bonomo23}.

\clearpage

\begin{deluxetable}{lll}
\label{tab:neid_data}
\tablecaption{NEID RVs of Kepler-21}
\tablehead{\colhead{BJD} & \colhead{RV (m s$^{-1}$)} & \colhead{$\sigma$ (m s$^{-1}$)}
}
\startdata
2459319.983253372 & -3.711289651058405 & 1.9977861994751576 \\
2459322.949158233 & 3.302805961702971 & 1.9918070377615071 \\
2459327.955391794 & 9.338296680380186 & 1.9935448130241864 \\
2459334.88414205 & 1.687730957375691 & 2.004277307640795 \\
2459343.9071232546 & 1.3653028848054545 & 2.010032152473358 \\
2459348.869819076 & 3.948535284095416 & 2.0251091007862194 \\
2459352.9566098084 & 1.7975607793018702 & 2.007984373563125 \\
2459374.8604099136 & 5.76822158926624 & 2.013228406883198 \\
2459384.932717996 & -5.612423105086472 & 2.0121043284307687 \\
2459390.914733203 & 9.379282343291242 & 2.012026298978136 \\
2459391.9571144907 & 0.1908261333323217 & 2.007995366785412 \\
2459394.726029862 & 8.366308987534328 & 2.6218713824862427 \\
2459407.754373812 & 3.5130256520259016 & 2.2092216962691675 \\
2459409.909124814 & 2.634606042709773 & 2.517624156488815 \\
2459413.8889704314 & 1.6332113607503511 & 4.713634420322408 \\
2459652.9739413573 & 2.70163058055367 & 2.034708281015946 \\
2459731.925124665 & -2.306563711602569 & 2.0636456188870307 \\
2459736.8921379047 & 0.6267107540249598 & 2.053504025403792 \\
2459737.9060579985 & -1.8811509443013128 & 2.054394137830588 \\
2459743.8339826358 & 4.689961862062039 & 2.0388508132975347 \\
\enddata
\end{deluxetable}

\begin{deluxetable}{lll}
\label{tab:hires_data}
\tablecaption{HIRES RVs of Kepler-21}
\tablehead{\colhead{BJD} & \colhead{RV (m s$^{-1}$)} & \colhead{$\sigma$ (m s$^{-1}$)}
}
\startdata
2455439.940758 & 7.764426254514014 & 1.0679568237296695 \\
2455440.8817225 & 1.7953107006725846 & 0.9036403891349272 \\
2455455.8286903333 & 5.062147638497088 & 1.20444500675834 \\
2455464.79112 & 18.36633499935585 & 1.2757189719519912 \\
2455465.8695983333 & 13.470272558194035 & 1.2268058904371675 \\
2455466.7280826666 & 6.153714793631367 & 1.2708426080298745 \\
2455467.8502383335 & 4.443947987900025 & 1.2165266725598365 \\
2455468.7165613333 & 6.815906051507717 & 1.2900972427316257 \\
2455469.755997 & 4.477680874110407 & 1.1536711650651779 \\
2455471.850213667 & 1.516578709377566 & 1.1498414277708588 \\
2455486.8226966667 & 6.392366708122006 & 1.255823531351599 \\
2455490.8218603334 & 3.0234692313439697 & 1.2493715394754608 \\
2455521.759713 & 3.6994367741342304 & 2.43083167076111 \\
2456476.001784 & -1.47679898411056 & 2.93147444725037 \\
2456487.060913 & 4.775301169994641 & 1.680384022736491 \\
2457241.066679 & 1.36382185553398 & 2.94111037254333 \\
2457294.9101505 & -2.5524544530966606 & 2.1813163336318704 \\
2457556.051661 & 8.17048804110913 & 3.21241617202759 \\
2458819.707526 & -16.3490133567182 & 2.86452221870422 \\
2460040.087155 & -24.2343090246205 & 2.72731447219849 \\
2460045.079269 & -15.293298811852502 & 2.52878427505493 \\
2460046.05493 & -10.5520872444347 & 2.67576360702515 \\
2460047.099471 & -20.2054354856567 & 2.8836829662323 \\
2460067.077338 & -17.2292798266838 & 2.57119631767273 \\
2460068.044293 & -24.9293287883024 & 2.3991322517395 \\
2460071.092932 & -13.0019737667089 & 2.75270795822144 \\
2460073.116579 & -16.5808002464388 & 2.47319889068604 \\
2460094.931953 & -20.0712133905834 & 2.69209456443787 \\
2460101.085875 & -11.8376109957227 & 2.55460834503174 \\
2460104.869354 & -6.57507877253908 & 2.4582850933075 \\
2460119.915898 & -14.095605739301599 & 2.45580315589905 \\
2460121.0312 & -10.3613629759099 & 2.32180786132812 \\
2460122.057863 & -5.25291236056075 & 2.30524373054504 \\
2460132.862741 & 2.94864336755811 & 2.5644588470459 \\
2460139.026302 & -15.183526697969201 & 2.75242590904236 \\
2460140.081655 & -2.51395044262316 & 2.54669785499573 \\
\enddata
\end{deluxetable}

\end{document}